\documentclass[conference,compsoc]{IEEEtran}
\IEEEoverridecommandlockouts
\ifCLASSOPTIONcompsoc
  \usepackage[nocompress]{cite}
\else
  \usepackage{cite}
\fi
\ifCLASSINFOpdf
\else
\fi
\usepackage[fleqn]{amsmath}
\usepackage{amssymb}
\usepackage{algorithm}
\usepackage[noend]{algpseudocode}

\usepackage{enumitem}
\newtheorem{theorem}{Theorem}
\newtheorem{lemma}{Lemma}
\usepackage{bm}
\usepackage{multirow}
\usepackage{booktabs}
\usepackage{hyperref}
\usepackage{color}
\usepackage[font=small,labelfont=bf]{caption}

\usepackage{tikz}
\usepackage{tikzpeople}
\usetikzlibrary{decorations.pathreplacing,calligraphy}
\usetikzlibrary{positioning,shadows,arrows,shapes.arrows,shapes.symbols,3d,calc,shapes}
\usepackage{pgfplots} 

\newcommand{\wzynote}[1]{\textcolor{red}{#1}}

\usepackage{ulem}

\begin{document}
%
\title{Bicoptor: Two-round Secure Three-party Non-linear Computation without Preprocessing for Privacy-preserving Machine Learning}




%
\author{\IEEEauthorblockN{Lijing Zhou\IEEEauthorrefmark{1},
Ziyu Wang$^{\ddagger}$\IEEEauthorrefmark{1}\thanks{$^{\ddagger}$ Ziyu Wang is the corresponding author.},
Hongrui Cui\IEEEauthorrefmark{2}, 
Qingrui Song\IEEEauthorrefmark{1} and
Yu Yu\IEEEauthorrefmark{2}}
\IEEEauthorblockA{\IEEEauthorrefmark{1}Huawei Technology, Shanghai, China\\
\{zhoulijing,wangziyu13,songqingrui1\}@huawei.com}
\IEEEauthorblockA{\IEEEauthorrefmark{2}Shanghai Jiao Tong University, Shanghai, China\\
\{rickfreeman,yyuu\}@sjtu.edu.cn}}



\maketitle

\begin{abstract}
The overhead of non-linear functions dominates the performance of the secure multiparty computation (MPC) based privacy-preserving machine learning (PPML).
This work introduces a family of novel secure three-party computation (3PC) protocols, Bicoptor, which improve the efficiency of evaluating non-linear functions. 
The basis of Bicoptor is a new sign determination protocol, which relies on a clever use of the truncation protocol proposed in SecureML (S\&P 2017). Our 3PC sign determination protocol only requires two communication rounds, and does not involve any preprocessing. 
Such sign determination protocol is well-suited for computing non-linear functions in PPML, e.g. the activation function ReLU, Maxpool, and their variants. We develop suitable protocols for these non-linear functions, which form a family of GPU-friendly protocols, Bicoptor.
All Bicoptor protocols only require two communication rounds without preprocessing.
We evaluate Bicoptor under a 3-party LAN network over a public cloud, and achieve more than 370,000 DReLU/ReLU or 41,000 Maxpool (find the maximum value of nine inputs) operations per second. 
Under the same settings and environment, our ReLU protocol has a one or even two orders of magnitude improvement to the state-of-the-art works, Falcon (PETS 2021) or Edabits (CRYPTO 2020), respectively without batch processing.
\end{abstract}

%
\IEEEpeerreviewmaketitle

\section*{Update}
In this updated version of our paper, which was originally presented at S\&P 2023~\cite{ZWC+-23}, we address certain security concerns raised by Xu et al~\cite{XLH24} in his paper regarding our DReLU protocol (Alg.~\ref{alg:drelu}). The concerns stem from an omission of a standard step: resharing, in our protocol. In MPC, resharing is free and default. In this latest version, we have included a detailed explanation of this aspect in App.~\ref{app:fix}.

\section{Introduction}
Secure multiparty computation (MPC)~\cite{Y82,GMW87,BGW88,B91,CCD88} is a fundamental cryptographic primitive that allows multiple parties to jointly evaluate any efficiently computable functions while preserving the input secrecy. An area that raises particular privacy concerns is machine learning (ML) where the predictive model is typically acquired by aggregating and analyzing sensitive data from numerous institutions. Moreover, performing inference operations on the model may also impose privacy concerns since mobile or IoT devices typically outsource sensitive data to cloud ML services.

Recently, MPC-based privacy-preserving machine learning (PPML), which strives to combine the utility of ML and the privacy-preserving guarantee of MPC, has received phenomenal attention from the research community. The key issue with this approach is the performance, since MPC would incur extra overhead on top of the considerably heavy ML operations. Various works aim to improve the performance of MPC PPML with different settings. Protocols such as Delphi~\cite{JVC18}, GAZZLE~\cite{MLS+-20}, CryptFlow2~\cite{RRK+-20}, ABY2.0~\cite{PSS+-21}, and Chameleon~\cite{RWT+-18} lie in two parties realm. SecureNN~\cite{WGC19}, Falcon~\cite{WTB+-21}, CryptGPU~\cite{TKT+21}, ABY3~\cite{MR18}, ASTRA~\cite{CCP+-19}, BLAZE~\cite{PS20}, and CryptFlow~\cite{KRC+-20} involve three parties. Fantastic~\cite{DEK21}, SWIFT~\cite{KPP+-21}, FLASH~\cite{BCP+-20}, and Trident~\cite{CRS20} are executed among four parties.

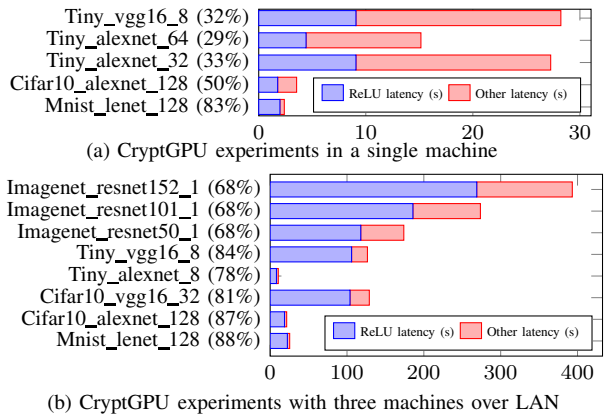
\begin{figure}[t]
    \centering
    \begin{tikzpicture}
      \begin{axis}[xbar stacked,
        legend pos = south east,
        legend style={font = \tiny},
        legend columns = 2,
        ytick={0,1,2,3,4},  
        yticklabels={
            Mnist\_lenet\_128\ (83\%),
            Cifar10\_alexnet\_128\ (50\%),
            Tiny\_alexnet\_32\ (33\%),
            Tiny\_alexnet\_64\ (29\%),
            Tiny\_vgg16\_8\ (32\%)
        },
        y tick label style={font = \footnotesize},
        xmin=0,
        height=3cm, 
        width=6cm, 
        bar width=0.2cm, 
        x tick label style={font = \footnotesize},
        x label style={at={(axis description cs:0.1, -0.15)},anchor=south, font=\footnotesize},
        xlabel={(a) CryptGPU experiments in a single machine},
        ]
        \addplot coordinates
        {(1.98,0) (1.77,1) (9.08,2) (4.42,3) (9.08,4) };
        \addplot coordinates
        {(2.39-1.98,0) (3.54-1.77,1) (27.27-9.08,2) (15.15-4.42,3) (28.22-9.08,4) };
        \legend{ReLU latency (s), Other latency (s)}
        \end{axis}
    \end{tikzpicture}
    \begin{tikzpicture}
      \begin{axis}[xbar stacked,
        legend pos = south east,
        legend style={font = \tiny},
        legend columns = 2,
        ytick={0,1,2,3,4,5,6,7},  
        yticklabels={
            Mnist\_lenet\_128\ (88\%),
            Cifar10\_alexnet\_128\ (87\%),
            Cifar10\_vgg16\_32\ (81\%),
            Tiny\_alexnet\_8\ (78\%),
            Tiny\_vgg16\_8\ (84\%),
            Imagenet\_resnet50\_1\ (68\%),
            Imagenet\_resnet101\_1\ (68\%),
            Imagenet\_resnet152\_1\ (68\%),
        },
        y tick label style={font = \footnotesize},
        xmin=0,
        height=4cm, 
        width=6cm, 
        bar width=0.2cm, 
        x label style={at={(axis description cs:0.1, -0.15)},anchor=south, font=\footnotesize},
        x tick label style={font = \footnotesize},
        xlabel={(b) CryptGPU experiments with three machines over LAN},
        ]
        \addplot coordinates
        {(22.393,0) (18.6645,1) (103.8843,2) (8.1937,3) (105.9647,4) (117.9001,5) (185.7585,6) (268.8527,7) };
        \addplot coordinates
        {(25.4026-22.393,0) (21.4969-18.6645,1) (129.0413-103.8843,2) (10.5601-8.1937,3) (126.5874-105.9647,4) (173.931-117.9001,5)
        (273.6108-185.7585,6) (393.2079-268.8527,7) };
        \legend{ReLU latency (s), Other latency (s)}
        \end{axis}
    \end{tikzpicture}
    \caption{The CryptGPU experiments are named by the dataset\_modelname\_batchsize, e.g., Tiny\_vgg16\_8 corresponds to the vgg16 model trained from the Tiny dataset, and the inference runs with a batch size of 8. The percentage after the name reflect the ratio of the ReLU latency to the total latency, e.g., ReLU spends 32\% of latency among the Tiny\_vgg16\_8 experiment in the local environment.}
    \label{fig:cryptgpu}
\end{figure}

Among published works, CryptGPU~\cite{TKT+21} represents the state-of-the-art. We deploy the CryptGPU implementation~\cite{TKT+-21-code} to run a sample PPML inference on both a single machine and a 3-party LAN network environment. The resulting runtimes are shown in Fig.~\ref{fig:cryptgpu}. We notice that under different network environments, the computation of Rectified Linear Unit (ReLU) takes a large portion of the overall runtime, ranging from around one-third (local) to three-quarters (LAN). This would be exacerbated in commercial deployment settings where WAN network offers an even worse network environment. 

The computation of non-linear functions (including ReLU) in CryptGPU is realized by the ABY3-based protocol~\cite{MR18} which is heavy in terms of communication overhead. In particular, the protocol for ReLU takes $3+\log_2\ell$ communication rounds and $45\ell$ bits of bandwidth, for a input $x \in \mathbb{Z}_q$ and $\ell:=\log_2q$. The ReLU function can be decomposed to Derivative ReLU (DReLU), i.e. a comparison between the input and zero (or determine the sign of the input), and a multiplication. We focus on DReLU since multiplication is a common task in MPC which already has highly optimized solutions. Currently, MPC-based comparison (CMP) protocols could be categorized into four types.
\begin{itemize}
  \item \textbf{A2B-CMP-B2A}: First switch input sharing from arithmetic form to binary form (A2B), then perform the bit-wise comparison to obtain the binary shares of the comparison result, and finally switch back to the arithmetic form (B2A)~\cite{MR18,KPP+-21,TKT+21}. Most notably, the ReLU protocol in CryptGPU utilizes this comparison method.
  \item \textbf{GC-based-CMP}: Directly apply the generic GC-based comparison protocol~\cite{MLS+-20,CCP+-19,PSS+-21,DSZ15,MZ17,CRS20}.
  \item \textbf{Random-masking-CMP}: Open the secret input masked by a random $r$, i.e., $e:=x+r$. Then the comparison of $x$ against any constant $c$ can be enabled by comparing $r$ against $e-c$. The comparison can be aided by additional auxiliary preprocessing information generated alongside $r$~\cite{WTB+-21,WGC19}. 
  \item \textbf{MSB-decomposition-CMP}: Decompose the input shares into the binary form and perform comparison~\cite{DSM+-20}.
\end{itemize}
Most of the works in the A2B-CMP-B2A, Random-masking-CMP, and MSB-decomposition-CMP categories require $\mathcal{O}(\log\ell)$ communication rounds ($\ell$ is the bit length of ring/field size). Despite that GC-based-CMP has a constant number of communication rounds, its bandwidth cost is usually the highest. Intuitively, communication is the bottleneck for all four comparison methods.

The performance of computing the Maxpool function is another bottleneck in PPML. In machine learning, it is usually required to determine the maximum element in a 2$\times$2 or 3$\times$3 matrix. Maxpool function involves several comparisons and dot products for element selections. Existing works in the literature mostly fall into two categories.
\begin{itemize}
	\item \textbf{Repeated comparison}: The greater of the first two elements are compared with the third element and so on~\cite{WTB+-21,WGC19}. 
	\item \textbf{Binary search}: Perform comparison in a binary tree manner, where the inputs are the leaf nodes and the maximum is at the root~\cite{DSM+-20}.
\end{itemize}
Like ReLU, protocols for Maxpool usually require dozens of communication rounds and Maxpool is another performance bottleneck for PPML.

We propose a novel and more efficient sign determination protocol, which implies a comparison protocol, to accelerate the overhead of non-linear functions like ReLU and Maxpool, in order to further optimize the performance of PPML. Our approach departs from existing methods, and improves the communication overhead compared to previous works. 

Our high-level idea for the comparison protocol is as follows. For an input $x\in[0,2^{\ell_x})\bigcup(q-2^{\ell_x},q)$, we define a variable $\xi=\xi(x):=x$ if $x\in[0,2^\ell_x)$ or $\xi=\xi(x):=q-x$ if $x\in(q-2^{\ell_x},q)$. $\xi$ can be recognized as the ``absolute value'' of $x$. We define the bit position of the most significant non-zero bit of $\xi$ as $\lambda-1$. We further define $\lambda$ as the effective bit length of $\xi$.~\footnote{For example, for $\xi=x=23=0b00010111$, $\ell_x=8$, $\lambda=5$, and $\xi_{\lambda-1}=\xi_4$.} The truncation protocol $\text{TRC}(x,k)$ represents truncating $k$-bits from $x$, which is originally proposed SecureML~\cite{MZ17}. Our idea is to study the outcome of $\text{TRC}(x,\lambda-1)$ or $\text{TRC}(x,\lambda)$. If the outcome is $1$ or $q-1$, then the input is positive or negative, respectively. Unfortunately, the truncation protocol proposed in SecureML~\cite{MZ17} may introduce a one-bit error, which poses a problem to our protocol. By carefully analyzing the behavior of the errors, we explicitly identify the conditions under which the errors occur (Lemma~\ref{lmm:truncation} and~\ref{lmm:truncation2}), and prove the inevitable existence of $1$ or $q-1$ even if errors occur.

Regarding the fact that the input value is unknown in an MPC context, and hence the $\lambda$ is unknown to all participants, we rely on performing repeated times of probabilistic truncations to compute the array $\{\text{TRC}(x,1),\cdots,\text{TRC}(x,\lambda-1),\text{TRC}(x,\lambda),\cdots,\text{TRC}(x,\ell_x)\}$, where $\ell_x$ is the precision of inputs. By using this method, we manage to construct a two-round 3-party sign determination protocol with better performance without relying on preprocessing.~\footnote{In general, MPC offline phase includes both preprocessing and distributing shared randomness. The overhead of distributing shared randomness (usually one time) is much cheaper than that of preprocessing. It is worth distinguishing between these two ideas for the rest of the paper. Most previous MPC-based PPML works~\cite{MR18,DSM+-20,TKT+21,DEK21,KPP+-21,PS20,BCP+-20,CRS20,WTB+-21} require preprocessing which is heavily computed.}

Based on the sign determination protocol, we further develop suitable 3PC protocols without preprocessing for common non-linear functions in PPML, e.g. ReLU, Maxpool, and their variants. It is worth mentioning that, the number of communication rounds of all our protocols is constant, i.e., 2, whereas the Maxpool protocols in Falcon~\cite{WTB+-21}, SecureNN~\cite{WGC19}, and CryptFlow~\cite{KRC+-20} require 104, 72, and 72 communication rounds in a typical setting ($n=9,\ell=40$) respectively.

\begin{table}[t]
  \caption{The comparison between the communication overhead of our ReLu/Maxpool protocol and that of other related works. (ss/gc: secret sharing/garble circuit. Comm.: the dominant one-pass communication cost is counted in bits. $\ell$: the bit length of $\mathbb{Z}_q=[0,q-1]$ ($\ell:=\log_2q$), e.g., $\ell = 40$. $p$: a prime field modulus, e.g., $p=67,\lceil \log_2p \rceil=7$. $\ell_x$: the precision of the input, e.g., $\ell_x=16$. $\kappa$: the computational security parameter, e.g., $\kappa=128$. $s$: the statistical security parameter, e.g., $s=40$. $n$: the number of inputs. 2R: 2-round. BT: binary tree.)}
  \label{tab:compare}
  \centering
    \linespread{1.25}
    \footnotesize
      \begin{tabular*}{8.5cm}{@{\extracolsep{\fill}} p{2.5cm} p{0.375cm} p{1.975cm} p{3.2cm} }
        \multicolumn{4}{c}{\textbf{ReLU}}\\
        \toprule
        \textbf{Protocol} & \textbf{Prep.} & \textbf{Round} & \textbf{Comm.(bit)} \\
        \hline
        ABY\_2PC\cite{DSZ15}          & Yes & 5              & $(2\kappa+20)\ell$ \\
        ABY2.0\_2PC\cite{PSS+-21}     & Yes & 4              & $(\kappa+3)\ell$ \\
        EMP\_2PC\cite{emp-toolkit}    & No  & 2              & $18\kappa\ell-6\kappa$ \\
        CryptFlow2\_2PC\cite{RRK+-20} & Yes & $4+\log\ell$   & $32(\ell+1) + 31\ell$  \\
        \hline
        Fantastic\_3PC\cite{DEK21}    & Yes & $3+\log\ell$   & $114\ell+6s+1$  \\
        BLAZE\_ss\_3PC\cite{PS20}     & Yes & $3+\log\ell$   & $16\ell$ \\
        BLAZE\_gc\_3PC\cite{PS20}     & Yes & $4$            & $(\kappa + 7)\ell$   \\
        SWIFT\_3PC\cite{KPP+-21}      & Yes & $3+\log\ell$   & $16\ell$ \\
        Falcon\_3PC\cite{WTB+-21}     & Yes & $5+\log\ell$   & $32\ell$  \\
        ABY3\_3PC\cite{MR18}          & Yes & $3+\log\ell$   & $45\ell$  \\
        CryptFlow\_3PC\cite{KRC+-20}  & No  & $10$           & $(6\log p + 19)\ell$  \\
        SecureNN\_3PC\cite{WGC19}     & No  & $10$           & $(8\log p + 24)\ell$  \\
        CryptGPU\_3PC\cite{TKT+21}    & Yes & $3+\log\ell$   & $45\ell$ \\
        Edabits\_3PC\cite{DSM+-20}    & Yes & $5+\log\ell$   & $80\ell$ \\
        \textbf{Ours\_3PC}            & \textbf{No} &\bm{$2$}& \bm{$(\ell_x + 2)\ell$}  \\		
        \hline
        Fantastic\_4PC\cite{DEK21}    & No  & $1+\log\ell$   & $44\ell+1$  \\
        SWIFT\_4PC\cite{KPP+-21}      & Yes & $1+\log\ell$   & $10\ell$ \\
        FLASH\_4PC\cite{BCP+-20}      & Yes & $2+\log\ell$   & $28\ell$  \\
        Trident\_4PC\cite{CRS20}      & Yes & $4$            & $8\ell + 4$ \\
        \bottomrule
        \multicolumn{4}{c}{\textbf{Maxpool}}\\
        \toprule
        \textbf{Protocol} & \textbf{Prep.} & \textbf{Round} & \textbf{Comm.(bit)} \\
        \hline
        SWIFT\_3PC\cite{KPP+-21}      & Yes & $\log n(3+\log\ell)$     & $(n-1)\cdot16\ell$ \\
        Falcon\_3PC\cite{WTB+-21}     & Yes & $(n-1)(7+\log\ell)$ & $(n-1)\cdot32\ell$  \\
        CryptFlow\_3PC\cite{KRC+-20}  & No  & $9(n-1)$                 & $(n-1)(6\log p + 19)\ell$  \\
        SecureNN\_3PC\cite{WGC19}     & No  & $10(n-1)$                & $(n-1)(8\log p + 24)\ell$  \\
        \textbf{Ours\_2R\_3PC}        & \textbf{No} &\bm{$2$}          &\bm{$\frac{n(n-1)}{2}(\ell_x+2)\ell$} \\
        \textbf{Ours\_BT\_3PC}        & \textbf{No} &\bm{$\log n$}     &\bm{$n(\ell_x+2)\ell$} \\
        \hline
        SWIFT\_4PC\cite{KPP+-21}      & Yes & $\log n(1+\log\ell)$     & $(n-1)\cdot16\ell$ \\
        \bottomrule
	  \end{tabular*}
\end{table}

\subsection{Related works}
In MPC-based PPML, the overhead of evaluating the non-linear functions, e.g., ReLU and Maxpool, dominates the total overhead. Existing protocols, such as ABY3~\cite{MR18}, Edabits~\cite{DSM+-20}, CryptGPU~\cite{TKT+21}, Fantastic~\cite{DEK21}, SWIFT~\cite{KPP+-21}, BLAZE~\cite{PS20}, FLASH~\cite{BCP+-20}, Trident~\cite{CRS20}, and Falcon~\cite{WTB+-21} mostly resort to preprocessing to enhance the online performance. In particular, after running an input-independent preprocessing phase which typically utilizes heavy cryptographic machinery, the parties are able to accomplish the PPML task relatively faster in the online phase once the inputs are ready. Notice that the total overhead (preprocessing and online) remains unchanged, i.e., with an improved performance for the online phase, the overhead of the preprocessing phase is usually heavy. For instance, Escudero et al.~\cite{DSM+-20} propose a comparison method where the online comparison performance could be improved by preprocessed material called ``Edabits''. The generation of Edabits relies on homomorphic encryption or oblivious transfer which incurs significant performance overhead. Our work aims to optimize the overall performance of different non-linear functions used in PPML.

As discussed above, the overhead of evaluating the ReLU and Maxpool functions accounts for a large portion of the total overhead of an inference in PPML, and the comparison operation is the core of ReLU and Maxpool. We thus review different secure comparison methods. Let the input $x\in\mathbb{Z}_q$ where $\log_2q=\ell$, the overheads of the four mainstream comparison protocol types are as follows. Note that ``the comparison between two secrets'', ``the sign determination'', and ``the comparison between a secret and a constant'' are equivalent in some way.
\begin{itemize}
	\item \textbf{A2B-CMP-B2A}:  ABY3~\cite{MR18}, SWIFT~\cite{KPP+-21}, and CryptGPU~\cite{TKT+21} first transform the secret input from arithmetic-form to Boolean form (A2B), then perform the bit-wise comparison, followed by a reversed transformation (B2A). This method usually takes $\mathcal{O}(\log \ell)$ rounds and communicates $\mathcal{O}(1)$ or $\mathcal{O}(\ell)$ bits in $\mathbb{Z}_q$. For example, the ReLU protocol in SWIFT~\cite{KPP+-21} takes $3+\log\ell$ rounds and $10\ell$ bits, 
	\item \textbf{GC-based-CMP}: GAZZLE~\cite{MLS+-20}, ASTRA~\cite{CCP+-19}, ABY2.0~\cite{PSS+-21}, ABY~\cite{DSZ15}, SecureML~\cite{MZ17}, and Trident~\cite{CRS20} apply the classical generic Yao garble-circuit (GC) method~\cite{Y82} to the secure comparison problem. Despite its optimized round overhead, the communication amount is usually significant. For instance, the communication bandwidth using EMP~\cite{emp-toolkit} is $18\kappa\ell-6\kappa=61,440$ bits under a typical setting of $\kappa=128,\ell=40$. 
	\item \textbf{Random-masking-CMP}: In Falcon~\cite{WTB+-21} and SecureNN~\cite{WGC19} the shares of an input $x$ is masked by a random $r$. The masked input $e:=x+r$ is then made public, and the result of comparing $r$ with $q/2-e$ thus reflects the relation between $x$ and $q/2$, which implies the sign of $x$. The overhead of this method is $\mathcal{O}(1)$ or $\mathcal{O}(\log\ell)$ rounds and $\mathcal{O}(\ell)$ bits, e.g, Falcon~\cite{WTB+-21}'s ReLU spends $5+\log\ell$ rounds and $32\ell$ bits, and SecureNN~\cite{WGC19}'s ReLU requires $10$ rounds and $8\log p + 24\ell$ bits, where $p$ is a small field modulus (e.g., $p=67$).
	\item \textbf{MSB-decomposition-CMP}: In Edabits~\cite{DSM+-20}, the shares of $x$, e.g., $[x]_0$ and $[x]_1$, is represented by $[x]_0 := [x]_0^{\prime\prime}\cdot \lceil \frac{q}{2} \rceil + [x]_0^\prime$ and $[x]_1 := [x]_1^{\prime\prime}\cdot \lceil \frac{q}{2} \rceil + [x]_1^\prime$. Then, checking whether $[x]_0^\prime + [x]_1^\prime\geq \lceil \frac{q}{2} \rceil$ and computing $[x]_0^{\prime\prime} \oplus [x]_1^{\prime\prime}$ offers the shares of two intermediate values $[\textsf{temp1}]$ and $[\textsf{temp2}]$, respectively. The sign of $x$ is obtained by $1-[\textsf{temp1}\oplus \textsf{temp2}]$. The Edabits~\cite{DSM+-20} communication overheads are $5+\log\ell$ rounds and $80\ell$ bits.
\end{itemize}
All protocols except the GC-based ones take more than two rounds of communication. Nevertheless, the GC-based method takes considerably more communication bandwidth compared to the secret-sharing-based ones. Thus our work enjoys a significant round complexity advantage compared to prior works.

For the maxpool layer in PPML, we investigate the performance of protocols for the functionality of finding the maximum element. 
\begin{itemize}
	\item \textbf{Repeated comparison}: The maximum protocols in CryptFlow~\cite{KRC+-20}, Falcon~\cite{WTB+-21}, and SecureNN~\cite{WGC19} get the output by sequentially comparing the output from the previous comparison to the next input element. Intuitively, the sequential operations take $\mathcal{O}(n)$ rounds. In particular, Falcon~\cite{WTB+-21}'s Maxpool requires $(n-1)\cdot(7+\log\ell)$ rounds and $(n-1)\cdot32\ell$ bits. 
	\item \textbf{Binary search}: SWIFT~\cite{KPP+-21} follows this method, in which the comparison is recursively applied to every different pair of inputs, until the output is narrowed down to the maximum. The protocol takes $\log n\cdot(3+\log\ell)$ rounds and $(n-1)\cdot16\ell$ bits.
\end{itemize}
Again, our Maxpool protocols outperform other protocols using these two methods in terms of round complexity. We compare the number of communication rounds and the cost of the dominant one-way bandwidth in our ReLU and Maxpool protocols against those in the literature in Tab.~\ref{tab:compare}.

\subsection{Our contributions}
We summarize the contributions of this work as follows.
\begin{itemize} [leftmargin=*]
	\item We define a variable $\xi=\xi(x):=x$ if an input $x\in[0,2^\ell_x)$ or $\xi=\xi(x):=q-x$ if $x\in(q-2^{\ell_x},q)$, and $\lambda$ refers to the effective bit length of $\xi$. We use $\text{TRC}(x,\lambda-1)$ or $\text{TRC}(x,\lambda)$ to represent truncating $\lambda$ or $\lambda-1$ bits from $x$. We prove the inevitable existence of $1$ or $q-1$ for a positive or a negative input even if errors occur (Lemma~\ref{lmm:oneexist1}). We further show the maximum numbers of truncations required to compute $\text{TRC}(x,\lambda-1)$ or $\text{TRC}(x,\lambda)$  in Lemma~\ref{lmm:oneexist2}. Based on Lemma~\ref{lmm:oneexist1} and Lemma~\ref{lmm:oneexist2}, we design a novel two-round 3PC DReLU protocol without preprocessing.
	\item After applying some optimization techniques, we further extend the DReLU protocol to other non-linear functions in PPML, including the Equality, ABS, ReLU, Dynamic ReLU (Leaky ReLU, PReLU, RReLU), ReLU6, Piecewise Linear Unit (PLU), MAX, MIN, SORT, and median (MED) functions. By carefully merging the multiplication operation(s) into the DReLU operation, we manage to achieve two rounds of communication for all the aforementioned protocols. In comparison, the corresponding ReLU or MAX protocols in prior works usually have dozens of rounds.
	\item We implement all the protocols and evaluate their performance under LAN settings (three VMs in the same cloud region). Without batching, our DReLU/ReLU protocols achieve a one or two orders of magnitude improvement compared with the state-of-the-art work Falcon~\cite{WTB+-21} or Edabits~\cite{DSM+-20}, respectively.~\footnote{We take the honest-majority and passive security settings. The ring size and the precision of inputs are set to 64 bits and 13 bits, respectively.} When the batch size is 100,000, we achieve 390,000 DReLU or 370,000 ReLU operations per second.
\end{itemize}
Most existing MPC protocols to evaluate non-linear functions rely on sequentially dependent computation, e.g., A2B switching or GC, and hence are not quite GPU-friendly. In comparison, all of our Bicoptor protocols (e.g., ReLU, CMP, and Maxpool) are suitable for GPU implementation since most of their computation steps are parallelizable.

For the rest of the paper, Sect.~\ref{sec:preliminary} introduces the notations, system settings, related backgrounds, and a key building block, i.e., the truncation protocol. Sect.~\ref{sec:drelu} intuitively presents the novel DReLU protocol. Sect.~\ref{sec:extension} extends DReLU to other non-linear functions in PPML. Sect.~\ref{sec:evaluate} analyses the concrete overhead of our protocols, and exhibits the performance of our implementations compared with Edabits~\cite{DSM+-20} and Falcon~\cite{WTB+-21}. Finally, Sect.~\ref{sec:conclude} concludes this paper. In Appendix, App.~\ref{app:fprn} introduces the fixed-point computation to handle decimal arithmetic. App.~\ref{app:functions} presents the non-linear functions used in this paper. App.~\ref{app:proofs} presents the proofs of all lemmas used in this paper. 
App.~\ref{app:dreluexp} proposes a concrete example for our DReLU protocol.

\begin{table}[t]
  \caption{Notation table.}
  \label{tab:notation}
  \centering
    \linespread{1.25}
    \footnotesize
      \begin{tabular*}{8.5cm}{@{\extracolsep{\fill}} l p{6.5cm} }
        \toprule
        \textbf{Notation} & \textbf{Description} \\
        \hline
        :=                          & defined as\\
        $i,j,k$                     & indexes\\
        $x$, $y$ or $\{x_i\}$       & a single input $x$ or $y$, or an input array $\{x_i\}$\\
        $\xi$                       & if an input $x \in [0, 2^{\ell_x})$ then $\xi=\xi(x):=x$;\\
                                    & if an input $x \in (q- 2^{\ell_x},q)$ then $\xi=\xi(x):=q-x$; \\
                                    & $\xi$ can be recognized as the ``absolute value'' of $x$;\\
                                    & the binary form of $\xi$ is $\{\xi_{\ell_x-1},\xi_{\ell_x-2},\cdots,\xi_{1},\xi_{0}\}$\\
        $q$, $\mathbb{Z}_q$         & an integer ring $\mathbb{Z}_q:=[0,q-1]$ with the prime modulus $q$\\
        $+,-,\cdot$                 & addition, subtraction, and multiplication in $\mathbb{Z}_q$ \\
        $\ell$, $\ell_x$            & $\ell:=\log_2q$; $\ell_x$ is the precision of the input\\
        $\lambda$                   & $\xi_{\lambda-1}$ is the most significant non-zero bit of $\xi$;\\
                                    & $\lambda$ is defined as the effective bit length of $\xi$\\
        $r$, $t$                    & a random mask, a random flipping bit \\
        $P_0$, $P_1$, $P_2$         & three participants \\
        $[x]$, $[x]_0$, $[x]_1$     & $[x] := ([x]_0, [x]_1)$ is the two-party secret sharing of $x$ \\
        $a,b,c,d,e$                 & a Beaver triple with $c=ab$, $d=x-a$, $e=y-b$ \\
        $\{u_i\},\{v_i\},\{w_i\}$   & arrays used in the DReLU protocol \\
        $\Pi\{\cdot\}$              & a random shuffle acting on an array \\
        $\vec{\bm{\phi}},\vec{\bm{\psi}},\vec{\bm{\theta}}$ & vectors used in the MAX/MIN/SORT/MED protocols \\
        $\bm{M}:=\{m_{i,j}\}$       & $m_{i,j}$ denotes the ($i,j$)-th element in the matrix $\bm M$ \\
        $x_i$, $\theta_i$           & $x_i$ or $\theta_i$ is the $i$-th item of $\{x_i\}$ or $\vec{\bm{\theta}}$ \\
        $[\{x_i\}]$, $[\vec{\bm{\theta}}]$ & the shares of the array $\{x_i\}$ and the vector $\vec{\bm{\theta}}$, resp. \\
        $\alpha,\beta,\gamma$       & constants \\
    \bottomrule
    \end{tabular*}
\end{table}

\section{Preliminary} \label{sec:preliminary}
In this section, we first introduce the system settings in Sect.~\ref{sub:sys}, including the topology, the security model, the definition of communication rounds, and the format of numbers used in this system. In Sect.~\ref{sub:secretsharing} we introduce the background of secret sharing. In Sect.~\ref{sub:truncation} we recall the truncation protocol proposed in SecureML~\cite{MZ17}. All notations used in this paper are listed in Tab.~\ref{tab:notation}.

\begin{figure}[t]
  \centering
  \begin{tikzpicture}[x=1cm,y=1cm,cap=round,align=center,
      fact/.style={rectangle, draw, rounded corners=1mm, fill=white, drop shadow,
            text centered, anchor=center, text=black},growth parent anchor=center,
      fact2/.style={rectangle, draw, rounded corners=1mm, fill=white,
            text centered, anchor=center, text=black},growth parent anchor=center]
  
  
      \node[rectangle, draw = black, minimum height = 2.25cm, minimum width = 8cm] at (0,-1.125) {}; 
    \node (u2) [alice, minimum size=0.5cm] at (0,-0.5) {};
      \node at (0,-1) [draw=none, anchor=center] {\small $P_2$};
      \node (u0) [dave, minimum size=0.5cm] at (-2.5,-1.5) {};
      \node at (-2.5,-2) [draw=none, anchor=center] {\small $P_0$};
      \node (u1) [bob, minimum size=0.5cm] at (2.5,-1.5) {};
      \node at (2.5,-2) [draw=none, anchor=center] {\small $P_1$};
      \node at (-4,-0.25) [draw=none, anchor=west] {\small Three participants};
  
      \node at (0.4,-0.3) [draw=none, anchor=west] {\small $\textsf{seed}_{02}$};
      \node at (0.4,-0.6) [draw=none, anchor=west] {\small $\textsf{seed}_{12}$};
  
      \node at (2.8,-1.3) [draw=none, anchor=west] {\small $\textsf{seed}_{01}$};
      \node at (2.8,-1.6) [draw=none, anchor=west] {\small $\textsf{seed}_{12}$};
  
      \node at (-4,-1.3) [draw=none, anchor=west] {\small $\textsf{seed}_{01}$};
      \node at (-4,-1.6) [draw=none, anchor=west] {\small $\textsf{seed}_{02}$};

      \node[rectangle, draw = black, minimum height = 0.5cm, minimum width = 8cm, anchor=west] at (-4,-3) {\small Client}; 
      \node at (-4,-2.5) [draw=none, anchor=west] {\small Secret sharing};
      \node at (2,-2.5) [draw=none, anchor=west] {\small Reconstruction};

    \draw [->] (-1,-2.75) -- (u0);
    \draw [->] (-1,-2.75) -- (u1);
  
    \draw [->] (u0) -- (0.95,-2.725);
    \draw [->] (u1) -- (1.05,-2.725);
  
    \draw [<->] (u2) -- (u0);
    \draw [<->] (u2) -- (u1);
    \draw [<->] (u1) -- (u0);

  \end{tikzpicture}
  \caption{System settings.}
  \label{fig:sys}
  \end{figure}
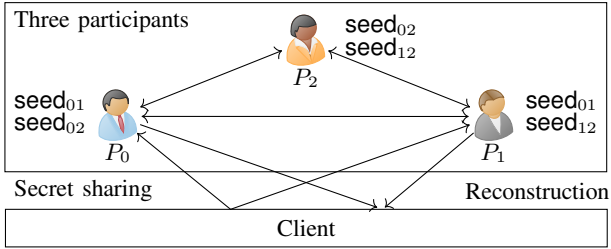

\subsection{System settings} \label{sub:sys}
A typical three-party computation (3PC) setting (Fig.~\ref{fig:sys}) is used in this paper, which is also commonly used in previous 3PC PPML works~\cite{DEK21,PS20,KPP+-21,WTB+-21,MR18,CCP+-19,KRC+-20,RWT+-18,WGC19,TKT+21}. The input(s) from a client is secretly shared between the participants. Then, the participants securely compute the corresponding shares of the result. Finally, the client reconstructs the shares of the result to form the output(s).

\noindent \textbf{Security model.} In our protocol, all three participants ($P_0$, $P_1$, and $P_2$) are static (i.e., non-adaptive) and honest-but-curious, i.e., semi-honest. We take the honest majority setting. It is assumed that there would be no collusion between any two of three participants. Hence, the protocol guarantees none of the participants can break the input, intermediate, or output secrecy alone.

\noindent \textbf{The number of communication rounds.}~\footnote{For a traditional two-party security protocol, a two-pass protocol is named as an ``one-round'' protocol. We do not use this definition in this paper.} In a 2PC or MPC protocol, some communication passes can be executed in parallel, and thus the round complexity of our protocol refers to the number of unparallelizable rounds, following the previous MPC-based PPML works~\cite{JVC18,MLS+-20,RRK+-20,PSS+-21,RWT+-18,WGC19,WTB+-21,TKT+21,MR18,CCP+-19,PS20,KRC+-20,DEK21,KPP+-21,BCP+-20,CRS20}.

\noindent \textbf{Pre-shared seed.} We assume that there are pre-shared pseudorandom seeds among participants, i.e., $P_0$ and $P_1$, $P_0$ and $P_2$, and $P_1$ and $P_2$, share the seeds $\textsf{seed}_{01}$, $\textsf{seed}_{02}$, and $\textsf{seed}_{12}$, respectively (as depicted in Fig.~\ref{fig:sys}). Note that the seeds should be kept secret from the other participant, e.g., $P_2$ does not know the value of $\textsf{seed}_{01}$.

\noindent \textbf{Complement Expression}. In a modern computer, a number is expressed in the complement form. For a positive number, the computer stores it in its original form; for a negative number, the complement code is used for storage. In this paper, we consider an integer ring $\mathbb{Z}_q=[0,q-1]$, in which $q$ is the modulus and the length of $q$ is $\ell: = \log_2q$. Unless explicitly stated, all arithmetic operations in this paper are assumed in $\mathbb{Z}_q$.

For a number $x$ of length $\ell_x$ in this ring, i.e., $x\in[0,2^{\ell_x})\bigcup(q-2^{\ell_x},q)$, it follows the constraint $\ell > \ell_x$. Take $q=2^{16},\ell_x = 8$ as an example. The positive integer $x=0b10111011$ is expressed as $0b0000000010111011$. The negation of $x$ is expressed as $q-x$, which is $0b1111111101010101$ in its binary form. For simplicity, we use integers to illustrate our protocols. A fixed-point computation method is introduced in App.~\ref{app:fprn} to handle decimal arithmetic.

\noindent \textbf{Non-linear Functions in PPML.} A function satisfying $F(x) = \alpha\cdot x + \beta$ is called a linear function. Otherwise, it is a non-linear function. Convolution is a typical example of linear functions. The common non-linear function used in machine learning is ReLU (Eq.~\ref{eq:relu}), which could be derived from DReLU (Eq.~\ref{eq:drelu}). The definitions for other non-linear functions are presented in App.~\ref{app:functions}.
\begin{equation} \label{eq:relu}
  \text{ReLU}(x) = \left\{ 
  \begin{aligned}
    &x,  \quad x\geq0 \quad \\
    &0,  \quad x<0  \quad \\
  \end{aligned}
  \right .
  = \text{DReLU}(x)\cdot x
\end{equation}
\begin{equation}  \label{eq:drelu}
\text{DReLU}(x) = \left\{
\begin{aligned}
	& 1, \quad x\geq0 \\
	& 0, \quad x<0 \\
\end{aligned}
\right .
\end{equation}

\subsection{Secret Sharing} \label{sub:secretsharing}
This paper focuses on the additive secret sharing scheme with an unbalanced setting. A plaintext message $x$ is shared between participants $P_0$ and $P_1$, which satisfies the relation $x := {[x]}_0 + {[x]}_1$. An unbalancing model means that $P_2$ does not hold any share of the input. Moreover, the shares have the linear homomorphic property, i.e., $[x] + \textsf{constant} = [x + \textsf{constant}]$, $[x_1] + [x_2] = [x_1 + x_2]$ and $\textsf{constant}\cdot[x] = [\textsf{constant}\cdot x]$.~\footnote{For $[x] + \textsf{constant}$, only one participant is required to add the constant to his secret share, e.g., $[x]_0 + \textsf{constant}$, while the other participant does nothing. Such notation is used to represent adding a constant number to a secret share for the rest of the paper.} If $x$ and $\textsf{constant}$ are both binary numbers, $[x] \oplus \textsf{constant}$ can be computed by $[x] + \textsf{constant} - 2\cdot \textsf{constant} \cdot [x]$. Similarly, $[x_1] \oplus [x_2] = [x_1] + [x_2] - 2\cdot [x_1 \cdot x_2]$ if both $x_1$ and $x_2$ are binary.

In order to perform multiplication among two shares, i.e., $[x\cdot y]$, Beaver triples~\cite{B91} (the shares of three correlated secrets $[a]$, $[b]$, and $[c]$) are utilized to decrease the online communication overhead. Participants first compute $[d] := [x - a]$ and $[e] := [y - b]$, then reconstruct $d$ and $e$. Next, the shares $[xy]$ could be locally obtained by $[x\cdot y] = d\cdot e + d\cdot[b] + e\cdot[a] + [c]$.~\footnote{The equation lies in $x\cdot y = (x-a+a)\cdot(y-b+b)=(d+a)(e+b)=(de+db+ea+ab)=(de+db+ea+c)$.} 

To further decrease the triple generation overhead, we use a trick proposed in~\cite{RWT+-18}. For more details, $P_0$ and $P_2$ generate $[a]_0$, $[b]_0$, and $[c]_0$ using $\textsf{seed}_{02}$, while $P_1$ and $P_2$ generate $[a]_1$ and $[b]_1$ using $\textsf{seed}_{12}$. Finally, $P_2$ computes a qualified $[c]_1 = ([a]_0 + [a]_1)\cdot ([b]_0 + [b]_1) - [c]_0$ and sends $[c]_1$ to $P_1$, which is the only communication overhead during the triple generation. This generation could be mirrored, in which $P_2$ computes $[c]_0$ and sends $[c]_0$ to $P_0$, correspondingly. If more than one triple is used in a protocol, the generation of half of the triples could be mirrored to balance the communication.

The seeds are also utilized to simplify the share generations for some special values. For example, when there is a need to have shares of $\textsf{constant}$ among $P_0$ and $P_1$. $P_0$ assigns a generated random ring element $r$ from $\textsf{seed}_{01}$ as $[\textsf{constant}]_0$, while $P_1$ regards $\textsf{constant} 
- r$ as $[\textsf{constant}]_1$. The zero-value and one-value secure shares would be utilized in our protocols.

\subsection{The Truncation Protocol with Errors} \label{sub:truncation}
SecureML~\cite{MZ17} proposes an MPC-based truncation protocol to keep the precision of a secretly shared fixed-point number during the computations. In this non-interactive protocol, $P_0$ and $P_1$ (respectively holding the ${[x]}_0$ and ${[x]}_1$ of a shared $x$) perform their own right shifting operation(s) on their shares individually. The $k$-bit non-cyclic right shifting is denoted by $\text{rShift}(x,k)$, without padding zero in the left hand. Specifically, $P_0$ directly right shifts its share for $k$ bits. $P_1$ takes the input negation and then does another negation after $k$-bit shifting. Finally, $P_0$ and $P_1$ withhold the shares
\begin{align*}
  [\text{TRC}(x, k)]_0 := &\text{rShift}([x]_0, k), \\
  [\text{TRC}(x, k)]_1 := &q - \text{rShift}(q - [x]_1, k).
\end{align*}

Due to the one-bit error, the $k$-bit truncation protocol for $x$ (denoted by $\text{TRC}(x, k)$) would lose one bit of accuracy at the least significant bit. We define the output of an exact truncation as that of $k$-bit right shifting, i.e., $\textsf{trc}:=\text{rShift}(x,k)$. Hence, it is possible that 
\begin{align*}
&[\text{TRC}(x, k)]_0 + [\text{TRC}(x, k)]_1\\
=&\text{TRC}(x, k) \in\{\textsf{trc}-1,\textsf{trc},\textsf{trc}+1 \}.
\end{align*}

For an $\ell_x$-precision input $x\in\mathbb{Z}_q$ and $\ell = \log_2{q}$, let $\xi=\xi(x):=x$ if $x \in [0, 2^{\ell_x})$ (a defined positive input and a zero input) and $\xi=\xi(x):=q-x$ if $x \in (q-2^{\ell_x},q)$ (a defined negative input). Hence, $\xi$ keeps as defined positive. Then, the truncation result with errors could be presented by Lemma~\ref{lmm:truncation}.
\begin{lemma} \label{lmm:truncation}
  In a ring $\mathbb{Z}_q$, let $x\in[0,2^{\ell_x})\bigcup(q-2^{\ell_x},q)$, where $\ell>\ell_x + 1$. Then we have the following results with probability $1-2^{\ell_x + 1 - \ell}$: 
  \begin{itemize}
      \item If $x\in [0,2^{\ell_x})$, then $\text{TRC}(x,k)=\text{rShift}(\xi,k)
 + \textsf{bit}$, where $\textsf{bit}=0$ or $1$.
      \item If $x\in (q-2^{\ell_x},q)$, then $\text{TRC}(x,k)=q-\text{rShift}(\xi,k) - \textsf{bit}$, where $\textsf{bit}=0$ or $1$.
  \end{itemize}
\end{lemma}
Lemma~\ref{lmm:truncation} proves the corresponding statement proposed in~\cite[Sect. 4.1]{MZ17} with a more precise characterization of the $\pm$1 error. We show the possible truncation error is $+1$ or $-1$ for a positive or a negative input respectively, while \cite[Theorem 1]{MZ17} simply states the existence of potential $\pm$1 error. Moreover, Lemma~\ref{lmm:truncation2} presents a special case of Lemma~\ref{lmm:truncation}, i.e., when errors do occur. As we will show later, our sign determination protocol benefits from a better understanding of how and when the $\pm$1 error occurs. The proofs of Lemma~\ref{lmm:truncation} and Lemma~\ref{lmm:truncation2} are presented in App.~\ref{app:proofs}.

\section{Two-round DReLU Protocol without Preprocessing} \label{sec:drelu}
In this section, we present the basis of Bicoptor, the sign determination (DReLU) protocol. We first describe how do we determine the sign of an input by using repeated truncations in Sect.~\ref{sub:repeated}. In Sect.~\ref{sub:strawmandrelu}, we utilize the result array from repeated truncations to form a ``strawman'' protocol which computes the DReLU function, but suffers from a few privacy issues. Finally, in Sect.~\ref{sub:formal} we show that the privacy issues can be solved by adding additional improvements on top of the strawman protocol, which leads to the complete DReLU protocol.
\subsection{The necessity of repeated truncations} \label{sub:repeated}
Recall that our aim is to determine the output of $\text{TRC}(x,\lambda-1)$ or $\text{TRC}(x,\lambda)$, where $\lambda$ is the effective bit-length of $\xi$. Lemma~\ref{lmm:oneexist1} proves that if $\text{TRC}(x,\lambda-1)$ or $\text{TRC}(x,\lambda)$ is $1$, the input is positive; if it is $q-1$, the input is negative. However, as mentioned before,  $\lambda$ is unknown to all participants in an MPC context. To address this problem, we can simply perform $\ell_x$ times of probabilistic truncations and output an array $\{\text{TRC}(x,1),\cdots,\text{TRC}(x,\lambda-1),\text{TRC}(x,\lambda),\cdots,\text{TRC}(x,\ell_x)\}$. Lemma~\ref{lmm:oneexist2} states that the result of $\text{TRC}(x,\lambda-1)$ and $\text{TRC}(x,\lambda)$ is included in this array. In other words, by checking the existence of $1$ or $q-1$ in this array, we can determine the sign of the input.

Lemma~\ref{lmm:tail1} and Lemma~\ref{lmm:tail2} prove that the behavior of the tail elements in the array follows a specific pattern. Based on Lemma~\ref{lmm:truncation}-\ref{lmm:tail2}, we formally introduce Theorem~\ref{thm:pattern}.
\begin{theorem} \label{thm:pattern}
  For an $\ell_x$-bits input $x\in\mathbb{Z}_q$, let $\xi=\xi(x):=x$ if $x \in (0, 2^{\ell_x})$, and let $\xi=\xi(x):=q-x$ if $x \in (q- 2^{\ell_x},q)$, in which $\ell:=\log_2q$. The binary form $\xi$ is defined as $\{\xi_{\ell_x-1},\xi_{\ell_x-2},\cdots,\xi_{1},\xi_{0}\}$, in which $\xi_i$ denotes the $i$-th bit and $\xi:=\sum_{i=0} ^{\ell_x-1} \xi_i\cdot 2^i$. $\lambda$ is the effective bit length of $\xi$, i.e., $\xi_{\lambda-1} = 1$ and $\lambda + 1 < \ell$. Set $\xi:=\xi^{\prime\prime}\cdot 2^k + \xi^\prime$, where $\xi^{\prime\prime}\in[0, 2^{\ell_x-k})$ and $\xi^\prime\in[0, 2^k)$, so that $\text{rShift}(\xi,k) = \xi^{\prime\prime}$. Then, for any value $\hat{\ell} \geq \lambda$, we have the following results with probability $1 - 2^{\lambda + 1 - \ell}$:
\begin{itemize}
  \item For $x = \xi$, there exists positive numbers $\lambda^\prime$ and $\lambda^{\prime\prime}$ ($\lambda^\prime \leq \lambda^{\prime\prime} \leq \ell_x$) satisfying $\text{TRC}(\xi, j) = 1$ for $\lambda^\prime \leq j \leq \lambda^{\prime\prime}$, and $\text{TRC}(\xi,j) = 0$ for $j > \lambda^{\prime\prime}$. 
  \item For $x = q - \xi$, there exists positive numbers $\lambda^\prime$ and $\lambda^{\prime\prime}$ ($\lambda^\prime \leq \lambda^{\prime\prime} \leq \ell_x$) satisfying $\text{TRC}(q - \xi, j) = q - 1$ for $\lambda^\prime \leq j \leq \lambda^{\prime\prime}$, and $\text{TRC}(q - \xi,j) = 0$ for $j > \lambda^{\prime\prime}$. 
\end{itemize}     
\end{theorem}

\begin{table}[t!]
  \caption{An example of truncations {\bf without} errors.}
  \label{tab:trunc1}
  \centering
    \linespread{1.25}
    \footnotesize
      \begin{tabular*}{8.5 cm}{@{\extracolsep{\fill}} l p{2.1cm} p{5.5cm}}
         & $x = 0b00010110$ & $x = q - 0b00010110$ \\
        \toprule
        \textbf{Opt.} & \textbf{Value}  & \textbf{Value} \\
        \hline
        TRC(x,1)  & $0b00001011$      & $q-0b00001011$ \\
        TRC(x,2)  & $0b00000101$      & $q-0b00000101$ \\
        TRC(x,3)  & $0b00000010$      & $q-0b00000010$ \\
        TRC(x,4)  & { $\color{red} \bf 0b00000001$} & { $\color{red} \bf q-0b00000001$} \\
        TRC(x,5)  & $0b00000000$      & $0b00000000$ \\
        TRC(x,6)  & $0b00000000$      & $0b00000000$ \\
        TRC(x,7)  & $0b00000000$      & $0b00000000$ \\
        TRC(x,8)  & $0b00000000$      & $0b00000000$ \\
        \bottomrule
  \end{tabular*}
\end{table}
\begin{table}[t!]
  \caption{An example of truncations {\bf with} errors.}
  \label{tab:trunc1-error}
  \centering
    \linespread{1.25}
    \footnotesize
      \begin{tabular*}{8.5 cm}{@{\extracolsep{\fill}} l p{2.2cm} p{0.5cm} p{2.2cm} p{0.5cm}}
         & $x = 0b00010110$   & & \multicolumn{2}{l}{$x = q - 0b00010110$}\\
        \toprule
        \textbf{Opt.} & \textbf{Value} & \textbf{Err.} & \textbf{Value} & \textbf{Err.} \\
        \hline
        TRC(x,1) & $0b00001011$   & $0$           & $q-0b00001011$  & $0$ \\          
        TRC(x,2) & $0b00000110$   & $+1$          & $q-0b00000110$  & $-1$\\           
        TRC(x,3) & $0b00000010$   & $0$           & $q-0b00000010$ & $0$ \\          
        TRC(x,4) &{$\color{red}\bf0b00000001$}&$0$& {$\color{red}\bf q-0b00000001$}& $0$\\
        TRC(x,5) &{$\color{red}\bf0b00000001$}&$+1$&{$\color{red}\bf q-0b00000001$}& $-1$ \\
        TRC(x,6) &{$\color{red}\bf0b00000001$}&$+1$&{$\color{red}\bf q-0b00000001$}& $-1$ \\
        TRC(x,7) & $0b00000000$   & $0$           & $0b00000000$  & $0$ \\
        TRC(x,8) & $0b00000000$   & $0$           & $0b00000000$  & $0$ \\
        \bottomrule
  \end{tabular*}
\end{table}
An example of the outcome array with $\ell_x$ times of exact truncations (without errors) for both a positive and a negative input is shown in Tab.~\ref{tab:trunc1}. Tab.~\ref{tab:trunc1-error} shows an example of the outcome array with $\ell_x$ times of probabilistic truncations (with errors). Due to the limited space, we postpone the formal proof to App.~\ref{app:proofs}.
\begin{table}[t!]
  \caption{Subtracting one from each element in the truncation result array (with errors).}
  \label{tab:trunc2}
  \centering
  \linespread{1.25}
  \footnotesize
\begin{tabular*}{8.5 cm}{@{\extracolsep{\fill}} l p{2.05cm} p{0.4cm} p{2.2cm} p{0.5cm}}
 & $x = 0b00010110$   & & \multicolumn{2}{l}{$x = q - 0b00010110$}\\
\toprule
\textbf{Opt.} & \textbf{Value} & \textbf{Err.} & \textbf{Value} & \textbf{Err.} \\
\hline
  $\text{TRC}(x,1)-1$ & $0b00001010$   & $0$           & $q-0b00001010$  & $0$ \\          
  $\text{TRC}(x,2)-1$ & $0b00000101$   & $+1$          & $q-0b00000101$  & $-1$\\           
  $\text{TRC}(x,3)-1$ & $0b00000001$   & $0$           & $q-0b00000001$ & $0$ \\ 
  $\text{TRC}(x,4)-1$  & { $\color{red} \bf 0b00000000$}& $0$ & $q-0b00000010$& $0$ \\
  $\text{TRC}(x,5)-1$  & { $\color{red} \bf 0b00000000$}& $+1$ & $q-0b00000010$& $-1$ \\
  $\text{TRC}(x,6)-1$  & { $\color{red} \bf 0b00000000$}& $+1$ & $q-0b00000010$& $-1$ \\
  $\text{TRC}(x,7)-1$  & $q-0b00000001$ & $0$   & $q-0b00000001$ & $0$\\
  $\text{TRC}(x,8)-1$  & $q-0b00000001$ & $0$   & $q-0b00000001$ & $0$\\
  \bottomrule
\end{tabular*}
\end{table}
\begin{algorithm}[t!]
	\hspace*{\algorithmicindent} \textbf{Input}: the shares of $x$ \\
	\hspace*{\algorithmicindent} \textbf{Output}: the shares of $\text{DReLU}(x)$ 
    \begin{algorithmic}[1]
    \Statex // \textit{$P_0$ and $P_1$ initialization.}
    \State $P_0$ and $P_1$ generate $\ell_x$ numbers of non-zero random ring elements $\{r_1,\cdots,r_{\ell_x}\}$ from $\textsf{seed}_{01}$.
    \State $P_0$ and $P_1$ set $[u_i] := [\text{TRC}(x,i)] - 1$ and $[v_i] := r_i \cdot [u_i]$, for $\forall i\in[1,\ell_x]$. 
    \State $P_0$ and $P_1$ set $[\{w_i\}]:=[\Pi\{v_i\}]$, using the shuffle-seed generated from $\textsf{seed}_{01}$.
    \State $P_0$ and $P_1$ send the shares $[\{w_i\}]$ to $P_2$.
    \Statex // \textit{$P_2$ processes.}
    \State $P_2$ reconstructs $\{w_i\}$, and sets $\text{DReLu}(x)=1$ if there exists zero(s) in the array; otherwise $\text{DReLu}(x) = 0$.
    \State $P_2$ shares $\text{DReLu}(x)$ to $P_0$ and $P_1$.
    \end{algorithmic}
  \caption{Strawman DReLU protocol.}
  \label{alg:strawmandrelu}
\end{algorithm}
\subsection{Strawman DReLU protocol} \label{sub:strawmandrelu}
After outputting an array $\{\text{TRC}(x,1),\cdots,\text{TRC}(x,\ell_x)\}$, we are now ready to produce the ``strawman'' protocol that computes the DReLU function. According to the secrecy requirement in our system, $P_2$ should not learn anything about the input $x$. Hence, $P_0$ and $P_1$ should mask and shuffle the shares of this array before revealing to $P_2$. The masking process could be achieved by multiplying each element in this array by a non-zero random ring element. The random ring elements and the shuffle-seed could be generated from a shared seed between $P_0$ and $P_1$, i.e., $\mathsf{seed}_{01}$. 

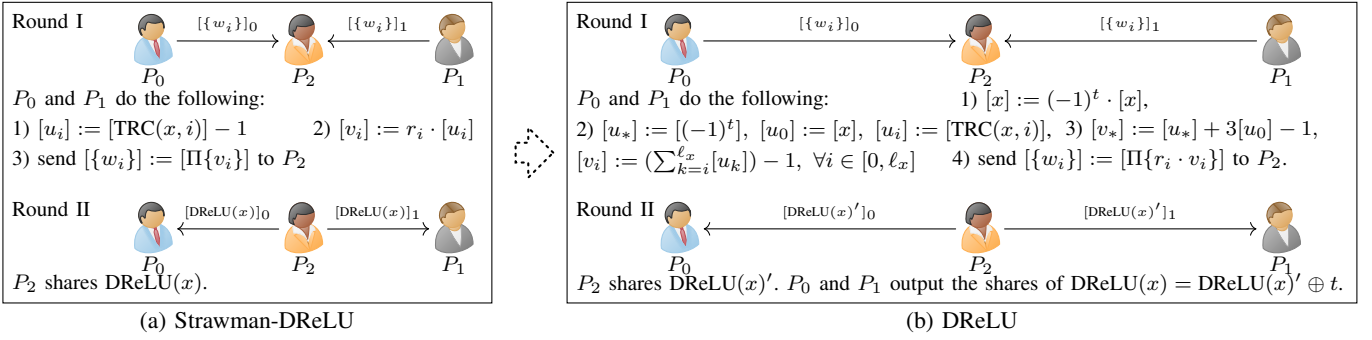
\begin{figure*}[t!]
  \centering
  \begin{tikzpicture}[x=1cm,y=1cm,cap=round,align=center,
      fact/.style={rectangle, draw, rounded corners=1mm, fill=white, drop shadow,
            text centered, anchor=center, text=black},growth parent anchor=center,
      fact2/.style={rectangle, draw, rounded corners=1mm, fill=white,
            text centered, anchor=center, text=black},growth parent anchor=center]
  
  
      \node[rectangle, draw = black, minimum height = 4cm, minimum width = 6.5cm] at (-5.75,-2) {}; 
      \node (u2) [alice, minimum size=0.5cm] at (-5,-0.5) {};
      \node at (-5,-1) [draw=none, anchor=center] {\footnotesize $P_2$};
      \node (u0) [dave, minimum size=0.5cm] at (-7,-0.5) {};
      \node at (-7,-1) [draw=none, anchor=center] {\footnotesize $P_0$};
      \node (u1) [bob, minimum size=0.5cm] at (-3,-0.5) {};
      \node at (-3,-1) [draw=none, anchor=center] {\footnotesize $P_1$};
      \node at (-9,-0.25) [draw=none, anchor=west] {\footnotesize Round I};
  
      \draw [<-] (u2) -- node [midway,above] {\tiny $[\{w_i\}]_0$} (u0);
      \draw [<-] (u2) -- node [midway,above] {\tiny $[\{w_i\}]_1$} (u1);

      \node at (-9,-1.3) [draw=none, anchor=west] {\footnotesize $P_0$ and $P_1$ do the following:};
      \node at (-9,-1.7) [draw=none, anchor=west] {\footnotesize 1) $[u_i]:=[\text{TRC}(x,i)]-1$};
      \node at (-5,-1.7) [draw=none, anchor=west] {\footnotesize 2) $[v_i]:=r_i\cdot [u_i]$};
      \node at (-9,-2.1) [draw=none, anchor=west] {\footnotesize 3) send $[\{w_i\}]:=[\Pi\{v_i\}]$ to $P_2$};

      \node [single arrow, draw, thick,dotted,fill=none,minimum width=0.3cm, minimum height=0.5cm, single arrow head extend=0.2cm,anchor=center] at (-2,-2) {}; 

      \node (uu2) [alice, minimum size=0.5cm] at (-5,-3) {};
      \node at (-5,-3.5) [draw=none, anchor=center] {\footnotesize $P_2$};
      \node (uu0) [dave, minimum size=0.5cm] at (-7,-3) {};
      \node at (-7,-3.5) [draw=none, anchor=center] {\footnotesize $P_0$};
      \node (uu1) [bob, minimum size=0.5cm] at (-3,-3) {};
      \node at (-3,-3.5) [draw=none, anchor=center] {\footnotesize $P_1$};
      \node at (-9,-2.75) [draw=none, anchor=west] {\footnotesize Round II};

      \draw [->] (uu2) -- node [midway,above] {\tiny $[\text{DReLU}(x)]_0$} (uu0);
      \draw [->] (uu2) -- node [midway,above] {\tiny $[\text{DReLU}(x)]_1$} (uu1);

      \node at (-9,-3.75) [draw=none, anchor=west] {\footnotesize $P_2$ shares $\text{DReLU}(x)$.};

      \node[rectangle, draw = black, minimum height = 4cm, minimum width = 10.5cm] at (3.75,-2) {}; 
      \node (u2) [alice, minimum size=0.5cm] at (4,-0.5) {};
      \node at (4,-1) [draw=none, anchor=center] {\footnotesize $P_2$};
      \node (u0) [dave, minimum size=0.5cm] at (0,-0.5) {};
      \node at (0,-1) [draw=none, anchor=center] {\footnotesize $P_0$};
      \node (u1) [bob, minimum size=0.5cm] at (8,-0.5) {};
      \node at (8,-1) [draw=none, anchor=center] {\footnotesize $P_1$};
      \node at (-1.5,-0.25) [draw=none, anchor=west] {\footnotesize Round I};

      \draw [<-] (u2) -- node [midway,above] {\tiny $[\{w_i\}]_0$} (u0);
      \draw [<-] (u2) -- node [midway,above] {\tiny $[\{w_i\}]_1$} (u1);

      \node at (-1.5,-1.3) [draw=none, anchor=west] {\footnotesize $P_0$ and $P_1$ do the following:};
      \node at (3.6,-1.3) [draw=none, anchor=west] {\footnotesize   1) $[x]:=(-1)^t\cdot[x]$,};
      \node at (-1.5,-1.7) [draw=none, anchor=west] {\footnotesize  2) $[u_*]:= [(-1)^t],\ [u_0]:=[x],\ [u_i]:=[\text{TRC}(x,i)]$,};
      \node at (5,-1.7) [draw=none, anchor=west] {\footnotesize     3) $[v_*]:= [u_*]+3[u_0]-1,$};
      \node at (-1.5,-2.1) [draw=none, anchor=west] {\footnotesize  $[v_i]:=(\sum_{k=i}^{\ell_x}[u_k]) - 1,\ \forall i\in[0,\ell_x]$};
      \node at (3.5,-2.1) [draw=none, anchor=west] {\footnotesize  4) send $[\{w_i\}]:=[\Pi\{r_i\cdot v_i\}]$ to $P_2$.};

      \node (uu2) [alice, minimum size=0.5cm] at (4,-3) {};
      \node at (4,-3.5) [draw=none, anchor=center] {\footnotesize $P_2$};
      \node (uu0) [dave, minimum size=0.5cm] at (0,-3) {};
      \node at (0,-3.5) [draw=none, anchor=center] {\footnotesize $P_0$};
      \node (uu1) [bob, minimum size=0.5cm] at (8,-3) {};
      \node at (8,-3.5) [draw=none, anchor=center] {\footnotesize $P_1$};
      \node at (-1.5,-2.75) [draw=none, anchor=west] {\footnotesize Round II};

      \draw [->] (uu2) -- node [midway,above] {\tiny $[\text{DReLU}(x)^\prime]_0$} (uu0);
      \draw [->] (uu2) -- node [midway,above] {\tiny $[\text{DReLU}(x)^\prime]_1$} (uu1);

      \node at (-1.5,-3.75) [draw=none, anchor=west] {\footnotesize $P_2$ shares $\text{DReLU}(x)^\prime$. $P_0$ and $P_1$ output the shares of $\text{DReLU}(x) = \text{DReLU}(x)^\prime\oplus t$.};

      \node at (-5.75,-4.25) [draw=none, anchor=center] {\small (a) Strawman-DReLU};
      \node at (3.75,-4.25) [draw=none, anchor=center] {\small (b) DReLU};
    
  \end{tikzpicture}
  \caption{Strawman-DReLU and DReLU protocol overview.}
  \label{fig:overview1}
  \end{figure*}

We immediately notice that, we are not able to determine the target element after multiplying $1$ or $q-1$ by a non-zero random ring element. To address this issue, we can either add one to or subtract one from each element in this array according to the needs. For simplicity, we use $\{u_i\}$ to represent the new array (Tab.~\ref{tab:trunc2}). By doing this, checking the existence of $1$ or $q-1$ in the original array is transferred into checking that of $0$ in $\{u_i\}$, which would not be affected by the masking process. Now, $P_2$ can reconstruct the shares from $P_0,P_1$ and output the result of the DReLU function according to the existence of $0$. We present the strawman DReLU protocol in Alg.~\ref{alg:strawmandrelu} and Fig.~\ref{fig:overview1}a. However, the strawman protocol still suffers from a few privacy issues.
\begin{itemize} [leftmargin=*]
	\item Despite multiplicative masking and shuffling, $P_2$ still learns the sign of the DReLU result which is not allowed according to our security definition.
	\item Due to the one-bit error, there might be continuous zeros in the array $\{u_i\}$ (Tab.~\ref{tab:trunc2}), which leaks information about the range of $x$.
	\item We omit the corner case in the strawman DReLU protocol. In particular, the protocol in Alg.~\ref{alg:strawmandrelu} does not work when $x=0$ or $x=1$.
\end{itemize}
In the next subsection, we show how to solve these problems using common techniques in secure multiparty computation.

\subsection{The DReLU procotol} \label{sub:formal}
In this subsection, we improve the strawman protocol to form the complete DReLU protocol.

\noindent \textbf{Masking the DReLU output with random input negation}. We can prevent $P_2$ from learning the actual sign of $x:={(-1)}^t\cdot x$ by flipping a coin $t$ using the shared $\textsf{seed}_{01}$ between $P_0$ and $P_1$. After $P_2$ responds $\text{DReLU}(x)^\prime$, $P_0$ and $P_1$ negate $\text{DReLU}(x)^\prime$ according to $t$. This negation $\text{DReLU}(x)=t\oplus \text{DReLU}(x)^\prime = t + \text{DReLU}(x)^\prime - 2t\cdot \text{DReLU}(x)^\prime$ is a linear operation since $P_0$ and $P_1$ knows $t$. Since $P_2$ does not know about $\textsf{seed}_{01}$, $\text{DReLU}(x)$ is completely masked by $t$. 

\noindent \textbf{Dealing with continuous zeros using summation}. To avoid the possible continuous zeros in the output array leaking information about the range of the input $x$, $P_0$ and $P_1$ could instead send the summation of subarrays, i.e., let $u_i:=\text{TRC}([x],i)$ and $v_i:=(\sum_{k=i}^{\ell_x}u_k) - 1$, for $i\in[1,\ell_x]$. By doing this, the continuous $1$-values in $\{u_i\}$ would result in a single $0$ in $\{v_i\}$ for a positive input. Tab.~\ref{tab:trunc3} depicts an example.

\noindent \textbf{Handling the input of one}. Notice that if the input value is $1$, then truncating it even by one position would result in zero instead of the $1$ or $q-1$ that we desire. Thus we append $[x]$ itself to the beginning of the output array $\{u_i\}$. This step is applied after the previous random negation operation.

\noindent \textbf{Handling the input of zero}. The DReLU function is defined to output $1$ at the input point of $0$, i.e., $\text{DReLU}(0):=1$. To accomplish this goal, we append an additional item to the truncation result array, denoted as $u_{*}$. In particular, we set $u_*:={(-1)}^t$ and $v_*:= u_*+{\sf coeff}\cdot u_0 - 1$ where $u_0$ should be $(-1)^{t} x$ (after the random negation). ${\sf coeff}$ is a positive constant. We want $v_*$ to satisfy the following constraint 
\begin{align*}
&\text{if and only if}\ x = 0, t = 0,\ \text{then}\ v_* = 0.
\end{align*}
This additional element $v_*$ assures the correct output of DReLU when $x=0$, regarding the flipping coin $t$, meanwhile, it does not affect the correctness of DReLU other inputs.

We select the value of $\sf coeff$ as $3$ and briefly discuss the reason to do so. Apparently when $x = 0$ the $\sf coeff$ term does not contribute to $v_*$ and thus we only need to concentrate on the $x \ne 0$ case, which should satisfy $(-1)^t (1 + {\sf coeff} \cdot x) \ne -1 \pmod{q}$. An intuitive way could be used to find the minimum $\sf coeff$ value, if the input $x$ ranges $(0,2^{\ell_x}) \bigcup (q-2^{\ell_x},q)$. When ${\sf coeff}=1$ or $2$, $x=q-2$ or $x=q-1$ breaks the aforementioned constraint, respectively. When ${\sf coeff}=3$, suppose the condition does not hold, which implies ${\sf coeff} \cdot x = 0$ or $-2 \pmod{q}$. Since we have $\ell_x \le \ell = \log_2 q$, it holds that ${\sf coeff} \cdot x < q$ if $x > 0$ and ${\sf coeff} \cdot x > -q$ if $x < 0$. Therefore the constraint that ${\sf coeff} \cdot x \ne 0$ and $\ne -2$ is satisfied.

Finally, these improvements are combined with our complete DReLU protocol (Alg.~\ref{alg:drelu} and Fig.~\ref{fig:overview1}b). $P_0$ and $P_1$ first generate a random bit $t$ from $\textsf{seed}_{01}$, set negate the input accordingly, i.e., $x:=(-1)^t\cdot x$. They also set $u_{*}$ and $u_0$ to deal with the special input zero and one. Then they execute a repeated times of probabilistic truncations on the input to get an array $\{u_i\}$, in which $u_i: = \text{TRC}(x,i),\forall i\in[1,\ell_x]$. Next, $v_*$ is set to $u_*+ 3 u_0-1$, and $v_i=(\sum_{k=i}^{\ell_x} u_k) - 1,\forall i\in[0,\ell_x]$. Recall the summation would eliminate the possible continuous zeros that leak input information. $P_0$ and $P_1$ mask and shuffle the output array using $\textsf{seed}_{01}$, i.e., $\{w_i\}:=\Pi\{r_i\cdot v_i\}$. $P_0$ and $P_1$ finally reshare and send $\{w_i\}$ to $P_2$. After $P_2$ reconstructs $\{w_i\}$ and sets $\text{DReLU}(x)^\prime$ according to whether $\{w_i\}$ contains zero, $P_2$ shares the output to $P_1$ and $P_0$ who subsequently removes the random negation to get shares of $\text{DReLU}(x)$. We demonstrate the processing of a positive input example in Fig.~\ref{fig:dreluexp} in App.~\ref{app:dreluexp}.

\begin{table}[t!]
  \caption{Summing-then-subtracting-one on the truncation results to avoid the leakage of the input range, {\bf with} one-bit errors.}
  \label{tab:trunc3}
  \centering
  \linespread{1.25}
  \footnotesize
\begin{tabular*}{8.5cm}{@{\extracolsep{\fill}} l p{2.5cm} p{3cm} }
  & $x = 0b00010110$ & $x = q - 0b00010110$ \\
  \toprule
  \textbf{Opt.} & \textbf{Value} & \textbf{Value} \\
  \hline
  $\sum_{k=1}^{8}\text{TRC}(x,k)-1$ & $0b00010101$      & $q-0b00010111$ \\
  $\sum_{k=2}^{8}\text{TRC}(x,k)-1$ & $0b00001010$      & $q-0b00001100$ \\
  $\sum_{k=3}^{8}\text{TRC}(x,k)-1$ & $0b00000100$      & $q-0b00000110$ \\
  $\sum_{k=4}^{8}\text{TRC}(x,k)-1$ & $0b00000010$      & $q-0b00000100$ \\
  $\sum_{k=5}^{8}\text{TRC}(x,k)-1$ & $0b00000001$      & $q-0b00000011$ \\
  $\sum_{k=6}^{8}\text{TRC}(x,k)-1$ & { $\color{red} \bf 0b00000000$} & $q-0b00000010$ \\
  $\sum_{k=7}^{8}\text{TRC}(x,k)-1$ & $q-0b00000001$      & $q-0b00000001$ \\
  $\sum_{k=8}^{8}\text{TRC}(x,k)-1$ & $q-0b00000001$      & $q-0b00000001$ \\
  \bottomrule
  \end{tabular*}
\end{table}
\begin{algorithm}[t!]
	\hspace*{\algorithmicindent} \textbf{Input}: the shares of $x$ \\
	\hspace*{\algorithmicindent} \textbf{Output}: the shares of $\text{DReLU}(x)$ 
    \begin{algorithmic}[1]
    \Statex // \textit{$P_0$ and $P_1$ initialization.}
    \State $P_0$ and $P_1$ generate $\ell_x + 2$ numbers of non-zero random ring elements $\{r_*,r_0,r_1,\cdots,r_{\ell_x}\}$ from $\textsf{seed}_{01}$.
    \label{alg:drelu:step:1}
    \State $P_0$ and $P_1$ set $[x]   := (-1)^t \cdot [x]$. \label{alg:drelu:step:reverse} 
    \State $P_0$ and $P_1$ set $[u_*] := [(-1)^t]$, $[u_0]:=[x]$, and $[u_i]:=[\text{TRC}(x,i)]$, $\forall i\in[1,\ell_x]$.
    \State $P_0$ and $P_1$ set $[v_*] := [u_*]+3\cdot [u_0] - 1$, and $[v_i] := (\sum_{k=i}^{\ell_x} [u_k]) - 1,\forall i\in[0,\ell_x]$.
    \State $P_0$ and $P_1$ set $[\{w_i\}]:=[\Pi\{r_i\cdot v_i\}]$, using the shuffle-seed generated from $\textsf{seed}_{01}$.
    \State $P_0$ and $P_1$ reshare~\footnotemark and send $[\{w_i\}]$ to $P_2$.
    \label{alg:drelu:step:6}
    \Statex // \textit{$P_2$ processes.}
    \State $P_2$ reconstructs $\{w_i\}$ and sets $\text{DReLu}(x)^\prime=1$ if there exists zero(s) in $\{w_i\}$; otherwise $\text{DReLu}(x)^\prime = 0$.
    \State $P_2$ shares $\text{DReLu}(x)^\prime$ to $P_0$ and $P_1$.
    \State $P_0$ and $P_1$ output the shares of $t\oplus \text{DReLu}(x)^\prime$. \label{alg:drelu:step:final} 
    \end{algorithmic}
  \caption{DReLU protocol.}
  \label{alg:drelu}
\end{algorithm}
\footnotetext{We acknowledge that the default resharing step was omitted in the version of our paper~\cite{ZWC+-23} presented at S\&P 2023; however, this common step was mentioned during our presentation at the conference (5:44 in \url{ https://www.youtube.com/watch?v=oaeycdolvVg}). This step addresses the security concerns regarding our protocol as raised by Xu et al.~\cite{XLH24}.}

\section{Protocols for Other Non-linear Functions} \label{sec:extension}
In this section, we describe how the DReLU protocol could be extended to securely compute various non-linear functions including the Equality, ABS, ReLU, Dynamic ReLU (Leaky ReLU, PReLU, RReLU), ReLU6, Piecewise Linear Unit (PLU), MAX, MIN, SORT, and median (MED) functions. The definition of these functions is presented in App.~\ref{app:functions}.

\subsection{DReLU Variants}
The protocols for the functions $\text{BitExt}(x)$, $\text{MSB}(x)$ and $\text{CMP}(x,y)$ could be derived from the $\text{DReLU}$ function (recall that we assume the inputs are encoded in the complementary form as in Sect.~\ref{sec:preliminary}).  We mainly introduce how to construct the Equality function in Alg.~\ref{alg:equality}, which invokes the DReLU protocol (Alg.~\ref{alg:drelu}) twice. For inputs $x$ and $y$, we have
\begin{align*}
    &\text{Equality}(x,y) =1 - \text{DReLU}(x-y)\oplus\text{DReLU}(y-x)\\
  = & 1 - (\text{DReLU}(x-y)^\prime \oplus t_0) \oplus (\text{DReLU}(y-x)^\prime \oplus t_1)\\
  = & 1 - \text{DReLU}^{\prime\prime} \oplus t^\prime,
\end{align*}
in which $t_0$ and $t_1$ are the random flipping bits in the two invocations, respectively. Specifically, after $P_0$ and $P_1$ invoke $\text{DReLU}(x-y)$ and $\text{DReLU}(y-x)$, the response from $P_2$ can be computed by an XOR operation, i.e., the shares of $\text{DReLU}^{\prime\prime}:=\text{DReLU}(x-y)^\prime\oplus \text{DReLU}(y-x)^\prime$. Finally, $P_0$ and $P_1$ also XOR the two random bits, i.e., $t^\prime:=t_0 \oplus t_1$, then obtain the shares of the Equality function output.
\begin{algorithm}[t]
	\hspace*{\algorithmicindent} \textbf{Input}: the shares of $x$ and $y$\\
	\hspace*{\algorithmicindent} \textbf{Output}: the shares of $\text{Equality}(x,y)$ 
    \begin{algorithmic}[1]
    \State $P_0$ and $P_1$ set $[x-y]=[x]-[y] $ and $[y-x]=[y]-[x] $ as the input shares of the first and second DReLU instance, using the random bits $t_0$ and $t_1$, respectively.
    \State $P_2$ obtains $\text{DReLU}(x-y)^\prime$ and $\text{DReLU}(y-x)^\prime$ from the two instances, sets $\text{DReLU}^{\prime\prime} := \text{DReLU}(x-y)^\prime \oplus \text{DReLU}(y-x)^\prime$, and shares $\text{DReLU}^{\prime\prime}$ to $P_0$ and $P_1$.
    \State $P_0$ and $P_1$ set $t^\prime:=t_0\oplus t_1$, and output the shares of $1 - t^\prime\oplus \text{DReLU}^{\prime\prime}$.
    \end{algorithmic}
  \caption{Equality protocol.}
  \label{alg:equality}
\end{algorithm}
\begin{algorithm}[t]
	\hspace*{\algorithmicindent} \textbf{Input}: the shares of $x$\\
	\hspace*{\algorithmicindent} \textbf{Output}: the shares of $\text{ReLU}(x)$ 
    \begin{algorithmic}[1]
    \State $P_0$ and $P_2$ generate $[a]_0,[b]_0,[c]_0$ from $\textsf{seed}_{02}$. $P_1$ and $P_2$ generate $[a]_1,[b]_1$ from $\textsf{seed}_{12}$
    \State $P_0$ and $P_1$ input $[x]$ to DReLU.
    \State $P_0$ sends $[d]_0: = [x]_0 - [a]_0$ to $P_1$, and $P_1$ sends $[d]_1: = [x]_1 - [a]_1$ to $P_0$. \label{alg:relu:step:rec}
	\State $P_2$ obtains $\text{DReLU}(x)^\prime$ from DReLU, computes 
  \Statex $[c]_1 := ([a]_0+[a]_1)\cdot([b]_0+[b]_1) - [c]_0$, and
  \Statex $e := \text{DReLU}(x)^\prime - ([b]_0+[b]_1)$; 
  \Statex and sends $e$ to $P_0$, $e$ and $[c]_1$ to $P_1$.
	\State $P_0$ and $P_1$ set $d = [d]_0 + [d]_1$, and computes the shares of the output, i.e., $[\text{ReLU}(x)]:= (1 - 2t)\cdot(d\cdot e + d[b] + e[a] + [c]) + t\cdot [x]$. \label{alg:relu:step:finish}
    \end{algorithmic}
  \caption{ReLU protocol.}
  \label{alg:relu}
\end{algorithm}

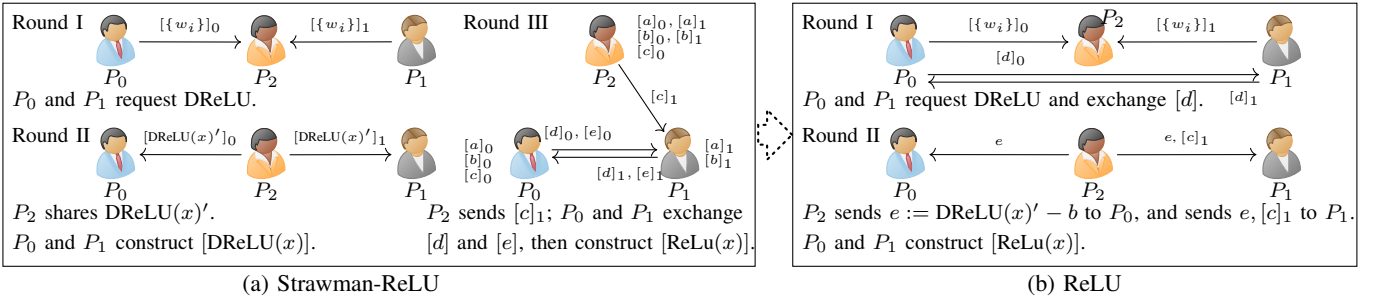
\begin{figure*}[ht!]
  \centering
  \begin{tikzpicture}[x=1cm,y=1cm,cap=round,align=center,
      fact/.style={rectangle, draw, rounded corners=1mm, fill=white, drop shadow,
            text centered, anchor=center, text=black},growth parent anchor=center,
      fact2/.style={rectangle, draw, rounded corners=1mm, fill=white,
            text centered, anchor=center, text=black},growth parent anchor=center]
  
  
      \node[rectangle, draw = black, minimum height = 3.5cm, minimum width = 10cm] at (-4,-1.75) {}; 
      \node (u2) [alice, minimum size=0.5cm] at (-5.5,-0.5) {};
      \node at (-5.5,-1) [draw=none, anchor=center] {\footnotesize $P_2$};
      \node (u0) [dave, minimum size=0.5cm] at (-7.5,-0.5) {};
      \node at (-7.5,-1) [draw=none, anchor=center] {\footnotesize $P_0$};
      \node (u1) [bob, minimum size=0.5cm] at (-3.5,-0.5) {};
      \node at (-3.5,-1) [draw=none, anchor=center] {\footnotesize $P_1$};
      \node at (-9,-0.25) [draw=none, anchor=west] {\footnotesize Round I};
  
      \draw [<-] (u2) -- node [midway,above] {\tiny $[\{w_i\}]_0$} (u0);
      \draw [<-] (u2) -- node [midway,above] {\tiny $[\{w_i\}]_1$} (u1);

      \node at (-9,-1.3) [draw=none, anchor=west] {\footnotesize $P_0$ and $P_1$ request DReLU.};

      \node [single arrow, draw, thick,dotted,fill=none,minimum width=0.3cm, minimum height=0.45cm, single arrow head extend=0.2cm,anchor=center] at (1.2,-1.75) {}; 

      \node (uu2) [alice, minimum size=0.5cm] at (-5.5,-2) {};
      \node at (-5.5,-2.5) [draw=none, anchor=center] {\footnotesize $P_2$};
      \node (uu0) [dave, minimum size=0.5cm] at (-7.5,-2) {};
      \node at (-7.5,-2.5) [draw=none, anchor=center] {\footnotesize $P_0$};
      \node (uu1) [bob, minimum size=0.5cm] at (-3.5,-2) {};
      \node at (-3.5,-2.5) [draw=none, anchor=center] {\footnotesize $P_1$};
      \node at (-9,-1.75) [draw=none, anchor=west] {\footnotesize Round II};

      \draw [->] (uu2) -- node [midway,above] {\tiny $[\text{DReLU}(x)^\prime]_0$} (uu0);
      \draw [->] (uu2) -- node [midway,above] {\tiny $[\text{DReLU}(x)^\prime]_1$} (uu1);

      \node at (-9,-2.8) [draw=none, anchor=west] {\footnotesize $P_2$ shares $\text{DReLU}(x)^\prime$.};
      \node at (-9,-3.2) [draw=none, anchor=west] {\footnotesize $P_0$ and $P_1$ construct $[\text{DReLU}(x)]$.};

      \node (uuu2) [alice, minimum size=0.5cm] at (-1,-0.5) {};
      \node at (-1,-1) [draw=none, anchor=center] {\footnotesize $P_2$};
      \node (uuu0) [dave, minimum size=0.5cm] at (-2,-2) {};
      \node at (-2,-2.5) [draw=none, anchor=center] {\footnotesize $P_0$};
      \node (uuu1) [bob, minimum size=0.5cm] at (0,-2) {};
      \node at (0,-2.5) [draw=none, anchor=center] {\footnotesize $P_1$};
      \node at (-3,-0.25) [draw=none, anchor=west] {\footnotesize Round III};

      \draw [->] (uuu2) -- (uuu1);
      \node at (-0.2,-1.25) [draw=none] {\tiny $[c]_1$};
      \draw [->] (-1.7, -1.95) -- node [near start,above] {\tiny $[d]_0,[e]_0$} (-0.3, -1.95);
      \draw [<-] (-1.7, -2.05) -- node [near end,below] {\tiny $[d]_1,[e]_1$} (-0.3, -2.05);

      \node at (-3.5,-2.8) [draw=none, anchor=west] {\footnotesize $P_2$ sends $[c]_1$; $P_0$ and $P_1$ exchange};
      \node at (-3.5,-3.2) [draw=none, anchor=west] {\footnotesize $[d]$ and $[e]$, then construct $[\text{ReLu}(x)]$.};

      \node at (-3,-1.9) [draw=none, anchor=west] {\tiny $[a]_0$};
      \node at (-3,-2.1) [draw=none, anchor=west] {\tiny $[b]_0$};
      \node at (-3,-2.3) [draw=none, anchor=west] {\tiny $[c]_0$};

      \node at (0.2,-1.9) [draw=none, anchor=west] {\tiny $[a]_1$};
      \node at (0.2,-2.1) [draw=none, anchor=west] {\tiny $[b]_1$};
      
      \node at (-0.7,-0.25) [draw=none, anchor=west] {\tiny $[a]_0,[a]_1$};
      \node at (-0.7,-0.45) [draw=none, anchor=west] {\tiny $[b]_0,[b]_1$};
      \node at (-0.7,-0.65) [draw=none, anchor=west] {\tiny $[c]_0$};

      \node[rectangle, draw = black, minimum height = 3.5cm, minimum width = 7.5cm] at (5.25,-1.75) {}; 
      \node (u2) [alice, minimum size=0.5cm] at (5.5,-0.5) {};
      \node at (5.75,-0.2) [draw=none, anchor=center] {\footnotesize $P_2$};
      \node (u0) [dave, minimum size=0.5cm] at (3,-0.5) {};
      \node at (3,-1) [draw=none, anchor=center] {\footnotesize $P_0$};
      \node (u1) [bob, minimum size=0.5cm] at (8,-0.5) {};
      \node at (8,-1) [draw=none, anchor=center] {\footnotesize $P_1$};
      \node at (1.5,-0.25) [draw=none, anchor=west] {\footnotesize Round I};

      \draw [<-] (u2) -- node [midway,above] {\tiny $[\{w_i\}]_0$} (u0);
      \draw [<-] (u2) -- node [midway,above] {\tiny $[\{w_i\}]_1$} (u1);

      \draw [->] (3.3, -0.95) -- node [near start,above] {\tiny $[d]_0$} (7.7, -0.95);
      \draw [<-] (3.3, -1.05) -- (7.7, -1.05); 
      \node at (7.5,-1.05) [draw=none, anchor=north] {\tiny $[d]_1$};

      \node at (1.5,-1.3) [draw=none, anchor=west] {\footnotesize $P_0$ and $P_1$ request DReLU and exchange $[d]$.};

      \node (uu2) [alice, minimum size=0.5cm] at (5.5,-2) {};
      \node at (5.5,-2.5) [draw=none, anchor=center] {\footnotesize $P_2$};
      \node (uu0) [dave, minimum size=0.5cm] at (3,-2) {};
      \node at (3,-2.5) [draw=none, anchor=center] {\footnotesize $P_0$};
      \node (uu1) [bob, minimum size=0.5cm] at (8,-2) {};
      \node at (8,-2.5) [draw=none, anchor=center] {\footnotesize $P_1$};
      \node at (1.5,-1.75) [draw=none, anchor=west] {\footnotesize Round II};

      \draw [->] (uu2) -- node [midway,above] {\tiny $e$} (uu0);
      \draw [->] (uu2) -- node [midway,above] {\tiny $e$, $[c]_1$} (uu1);

      \node at (1.5,-2.8) [draw=none, anchor=west] {\footnotesize $P_2$ sends $e:=\text{DReLU}(x)^\prime-b$ to $P_0$, and sends $e,[c]_1$ to $P_1$.}; 
      \node at (1.5,-3.2) [draw=none, anchor=west] {\footnotesize $P_0$ and $P_1$ construct $[\text{ReLu}(x)]$.};

      \node at (-4.5,-3.75) [draw=none, anchor=center] {\small (a) Strawman-ReLU};
      \node at (5.25,-3.75) [draw=none, anchor=center] {\small (b) ReLU};
    
  \end{tikzpicture}
  \caption{Strawman-ReLU and ReLU protocol overview.}
  \label{fig:overview2}
  \end{figure*}

\subsection{ReLU}
As mentioned in Sect.~\ref{sub:sys}, the ReLU function could be decomposed into DReLU multiplied by the input value. Trivially, $P_0$ and $P_1$ could utilize the Beaver triple to perform this multiplication on shared values, as shown in  Fig.~\ref{fig:overview2}a. Nevertheless, when the triple is generated by $P_2$, the communication for DReLU and triple generation could be integrated to further reduce the round number from three to two, as shown in Fig.~\ref{fig:overview2}b. In particular, notice that after completing the DReLU protocol, $P_2$ holds $\text{DReLU}(x)^\prime$, and 
$$\text{DRelu}(x)=t\oplus \text{DRelu}(x)^\prime = t + \text{DRelu}(x)^\prime - 2\cdot t\cdot \text{DRelu}(x)^\prime.$$
Moreover since $\text{ReLU}(x) = x\cdot \text{DReLU}(x)$, we have
\begin{align*}
  &[\text{ReLU}(x)] = (1-2t)\cdot [x \cdot \text{DRelu}(x)^\prime ] + t\cdot [x]\\
 =&(1 - 2t)\cdot(d\cdot e + d[b] + e[a] + [c]) + t\cdot [x].
\end{align*}
The share multiplication is transformed from $[\text{DRelu}(x) \cdot x]$ to $[\text{DRelu}(x)^\prime \cdot x]$, which is the key to the communication optimization. It is worth noticing that $e := \text{DRelu}(x)^\prime - b$ does not break the secrecy of $\text{DRelu}(x)^\prime$, since none of $P_0$ and $P_1$ knows the integrate $b$ value. In this way, we manage to decrease the communication rounds required for the protocol down to two. This process is shown in Alg.~\ref{alg:relu}

\subsection{ABS, Dynamic ReLU, MAX2, MIN2, and Funnel ReLU} \label{sub:reluextend}
Building on the ReLU protocol, the protocols for the ABS (Eq.~\ref{eq:abs}), Dynamic Relu (Eq.~\ref{eq:dynamicrelu}), MAX2 (Eq.~\ref{eq:max}), MIN2 (Eq.~\ref{eq:min}), and Funnel ReLU (Eq.~\ref{eq:funnelrelu}) functions could be constructed as follows. 
\\
\newline
\noindent \textbf{ABS}. Since $\text{ABS}(x) =  (2\cdot\text{DReLU}(x) - 1)\cdot x$,
\begin{align*}
&[\text{ABS}(x)] =  [(2\cdot\text{DReLU}(x) - 1)\cdot x] \\
= & [(2\cdot(\text{DReLU}(x)^\prime \oplus t) - 1)\cdot x] \\
= & [(2\cdot(\text{DReLU}(x)^\prime + t - 2\cdot \text{DReLU}(x)^\prime \cdot t) - 1)\cdot x] \\
= & (2-4t)\cdot[x \cdot \text{DReLU}(x)^\prime ] + (2t-1)\cdot [x].
\end{align*}
Hence, $P_0$ and $P_1$ would computes
$$(2-4t)\cdot(de+d[b]+e[a]+c) + (2t-1)\cdot [x]$$
in the step~\ref{alg:relu:step:finish} of Alg.~\ref{alg:relu} for the ABS function. 
\\
\newline
\noindent \textbf{Dynamic ReLU}. Similarly, $P_0$ and $P_1$ would computes
\begin{align*}
&(1-2t)\cdot(\alpha_1-\alpha_0)\cdot(de+d[b]+e[a]+c) \\
+& ((\alpha_1-\alpha_0)t+\alpha_0)\cdot [x]
\end{align*}
in the step~\ref{alg:relu:step:finish} of Alg.~\ref{alg:relu} for the Dynamic ReLU function.
\\
\newline
\noindent \textbf{MAX2 and MIN2}. Moreover, for $\text{MAX2}(x,y)$ and $\text{MIN2}(x,y)$, $P_0$ and $P_1$ would reconstruct $d:=(x-y)-a$ or $d:=(y-x)-a$ in the step~\ref{alg:relu:step:rec}, then compute
$$(1-2t)\cdot(de+d[b]+e[a]+c) + t\cdot [x-y] + [y],$$
$$\text{or}\ (1-2t)\cdot(de+d[b]+e[a]+c) + t\cdot [y-x] + [x]$$
in the step~\ref{alg:relu:step:finish} of Alg.~\ref{alg:relu}, respectively.
\\
\newline
\noindent \textbf{Funnel ReLU}. For the Funnel ReLU function, the reconstructed $d$ should be $(x-\text{T}(x))-a$ in the step~\ref{alg:relu:step:rec}, and the computation in the step~\ref{alg:relu:step:finish} of Alg.~\ref{alg:relu} is
$$(1-2t)\cdot(de+d[b]+e[a]+c) + t\cdot [x-\text{T}(x)] + [\text{T}(x)].$$
Since we only modify the message contents used in the two-round ReLU protocol in Alg.~\ref{alg:relu}, all the protocols described here keep the round complexity of two.

\subsection{Piecewise Linear Unit (PLU)}
The ReLU and other variant functions we considered above could be viewed as piecewise functions with two segments, which involve one DReLU. More generally, the number of DReLU invocations is equal to the number of segments in a piecewise function minus one. Moreover, for $m+2$ segments in a PLU function (Eq.~\ref{eq:plu}), there would be $m+1$ interval points (i.e., $\gamma_j,\forall j\in[0,m]$) and $m+2$ linear function coefficients (i.e., $\alpha_j,\beta_j,\forall j\in[0,m+1]$), there would be $m+1$ DReLU invocations and $m+2$ multiplications.

For the $j$-th segment, the corresponding monomial is
\begin{align*}
&(\text{DReLU}(x-\gamma_{j-1}) \oplus \text{DReLU}(x-\gamma_j))\cdot (\alpha_j\cdot x + \beta_j)\\
=&(t_{j-1}\oplus\text{DReLU}(x-\gamma_{j-1})^\prime \oplus t_j\oplus \text{DReLU}(x-\gamma_j)^\prime)\\
&\cdot (\alpha_j\cdot x + \beta_j)\\
=&(t_{j-1,j}^\prime\oplus\text{DReLU}_{j-1,j}^{\prime\prime})\cdot (\alpha_j\cdot x + \beta_j)\\
=&\alpha_j\cdot(1-2t_{j-1,j}^\prime)\cdot\text{DReLU}_{j-1,j}^{\prime\prime}\cdot x + \alpha_j\cdot t_{j-1,j}^\prime \cdot x \\
&+ \beta_j \cdot (1-2t_{j-1,j}^\prime)\cdot \text{DReLU}_{j-1,j}^{\prime\prime} + \beta_j \cdot t_{j-1,j}^\prime,
\end{align*}
where $P_0$ and $P_1$ computes $t_{j-1,j}^\prime:=t_{j-1}\oplus t_j$ and $P_2$ computes $\text{DReLU}_{j-1,j}^{\prime\prime}:=\text{DReLU}(x-\gamma_{j-1})^\prime\oplus\text{DReLU}(x-\gamma_j)^\prime$. The two special end-point segments (the $m$-th and $0$-th ones) are as follows.
\begin{align*}
&(\text{DReLU}(x-\gamma_m) \oplus 0)\cdot (\alpha_{m+1}\cdot x + \beta_{m+1})\\
=&\text{DReLU}(x-\gamma_m) \cdot (\alpha_{m+1}\cdot x + \beta_{m+1})\\
=&(t_m\oplus\text{DReLU}(x-\gamma_m)^\prime)\cdot (\alpha_{m+1}\cdot x + \beta_{m+1})\\
=&\alpha_{m+1}\cdot(1-2t_m)\cdot\text{DReLU}(x-\gamma_m)^\prime\cdot x + \\
&+ \beta_{m+1} \cdot (1-2t_m)\cdot \text{DReLU}(x-\gamma_m)^\prime + \\
&+ \alpha_{m+1}\cdot t_m \cdot x + \beta_{m+1} \cdot t_m,
\end{align*}
and
\begin{align*}
&(1\oplus \text{DReLU}(x-\gamma_0))\cdot (\alpha_0\cdot x + \beta_0)\\
=&(1\oplus t_0\oplus\text{DReLU}(x-\gamma_0)^\prime)\cdot (\alpha_0\cdot x + \beta_0)\\
=&\alpha_0\cdot(2t_0-1)\cdot\text{DReLU}(x-\gamma_0)^\prime\cdot x + \alpha_0\cdot (1-t_0) \cdot x \\
&+ \beta_0 \cdot (2t_0-1)\cdot \text{DReLU}(x-\gamma_0)^\prime + \beta_0 \cdot (1-t_0).
\end{align*}

\begin{algorithm}[t]
	\hspace*{\algorithmicindent} \textbf{Input}: the shares of $x$\\
	\hspace*{\algorithmicindent} \textbf{Output}: the shares of $\text{PLU}(x)$ 
    \begin{algorithmic}[1]
    \State $P_0$ and $P_1$ generate the shares of $m+2$ numbers of triples from $\textsf{seed}_{02}$ and $\textsf{seed}_{12}$, respectively, where all $a$ values are identical. Skip the $j$-th triple if $\alpha_j = 0, \forall j \in [0,m+1]$.
    \State $P_0$ sends $[d]_0: = [x]_0 - [a]_0$ to $P_1$, and $P_1$ sends $[d]_1: = [x]_1 - [a]_1$ to $P_0$. 
    \State $P_0$ and $P_1$ input $[x-\gamma_j]$ to the $j$-th DReLU instance, $\forall j \in [0,m]$.
    \State $P_2$ obtains $\text{DReLU}(x-\gamma_j)^\prime$ from the $j$-th DReLU instance, and computes $\text{DReLU}_{j-1,j}^{\prime\prime}:=\text{DReLU}(x-\gamma_{j-1})^\prime\oplus\text{DReLU}(x-\gamma_j)^\prime$, $\forall j\in[0,m]$.
    \State $P_2$ sets $e_0:=\text{DReLU}(x-\gamma_0)^\prime - b_0$, $e_j:=\text{DReLU}_{j-1,j}^{\prime\prime} - b_j$ ($\forall j\in[1,m]$), $e_{m+1}:=\text{DReLU}(x-\gamma_{m})^\prime - b_{m+1}$.
    \State $P_2$ sends $e_j$ to $P_0$ and $e_j,[c_j]_1$ to $P_1$, $\forall j\in[0,m+1]$.
    \State $P_0$ and $P_1$ reconstruct $d$, set $t_{j-1,j}^\prime:=t_{j-1}\oplus t_j$, $\forall j\in[1,m]$, and output the shares
      \begin{align*}
      =&\alpha_{m+1}\cdot(1-2t_m)\\
      &\cdot(de_{m+1}+d[b_{m+1}]+e_{m+1}[a]+[c_{m+1}]) \\
      &+ \beta_{m+1} \cdot (1-2t_m)\cdot \text{DReLU}(x-\gamma_m)^\prime  \\
      &+ \alpha_{m+1}\cdot t_m \cdot x + \beta_{m+1} \cdot t_m \\
      +&... \\
      +&\alpha_j\cdot(1-2t_{j-1,j}^\prime)\cdot(de_j+d[b_j]+e_j[a]+[c_j]) \\
      &+ \beta_j \cdot (1-2t_{j-1,j}^\prime)\cdot \text{DReLU}_{j-1,j}^{\prime\prime}  \\
      &+ \alpha_j\cdot t_{j-1,j}^\prime \cdot x + \beta_j \cdot t_{j-1,j}^\prime\\
      +&...\\
      +&\alpha_0\cdot(2t_0-1)\cdot(de_0+d[b_0]+e_0[a]+[c_0]) \\
      &+ \beta_0 \cdot (2t_0-1)\cdot \text{DReLU}(x-\gamma_0)^\prime \\
      &+ \alpha_0\cdot (1-t_0) \cdot x   + \beta_0 \cdot (1-t_0),
      \end{align*}
    \end{algorithmic}
  \caption{PLU protocol.}
  \label{alg:plu}
\end{algorithm}

The shares of the triples ($[a_j],[b_j],[c_j], \forall j\in[0,m+1]$) are used to compute the shares $[\text{DReLu}(x-\gamma_0)^\prime\cdot x]$, $[\text{DReLU}_{j-1,j}^{\prime\prime}\cdot x]$ ($\forall j\in[1,m]$), and $[\text{DReLu}(x-\gamma_m)^\prime\cdot x]$. In total, there would be $m+2$ numbers of triples being consumed. Note that the multiplicand $x$ is the same in these $m+2$ numbers of multiplications (if all $\alpha_j$ are not zero), we would use the same $a$ value in these triples, leading to the same $d:=x-a$ value. Alg.~\ref{alg:plu} describes the protocol in detail. For simplicity, we consider a non-mirror case in Alg.~\ref{alg:plu}. As we discussed in Sect.~\ref{sub:secretsharing}, $\lceil \frac{m+2}{2} \rceil$ triples could be generated regularly (i.e., $P_2$ sends $[c]_1$ to $P_1$) while the generation of the remained $\lfloor \frac{m+2}{2} \rfloor$ triples could be mirrored (i.e., $P_2$ sends $[c]_0$ to $P_0$). We achieve this balancing in our implementation.

ReLU6 is a special case for PLU, in which $m=1,\alpha_0=\beta_0=\beta_1=\alpha_2=\gamma_0=0,\alpha_1=1,\beta_2=\gamma_2=6$.
\begin{figure*}[t!]
\centering
\begin{tikzpicture}[x=1cm,y=1cm,cap=round,align=center,
    fact/.style={rectangle, draw, rounded corners=1mm, fill=white, drop shadow,
          text centered, anchor=center, text=black},growth parent anchor=center,
    fact2/.style={rectangle, draw, rounded corners=1mm, fill=white,
          text centered, anchor=center, text=black},growth parent anchor=center]


	\node[fact, rectangle, draw = black, minimum height = 3cm, minimum width = 11.5cm, anchor = west] at (-9, -2) {}; 
    
	\node[fact, rectangle, draw = black, minimum height = 3.6cm, minimum width = 4.2cm, anchor = west] at (4.8, -2.25) {}; 

    \node at (-9,-0.25) [draw=none, anchor=west] {\small Input: {12, 46, 31, 27}}; 

	\node[fact2, rectangle, draw = black, minimum height = 2.25cm, minimum width = 1cm, anchor = west] at (-8.7,-2.1) {}; 

	\node[fact2, rectangle, draw = black, minimum height = 2.25cm, minimum width = 1cm, anchor = west] at (-6.7,-2.1) {}; 

	\node[fact2, rectangle, draw = black, minimum height = 2.25cm, minimum width = 7cm, anchor = west] at (-4.7,-2.1) {}; 

	\node[fact2, rectangle, draw = black, minimum height = 2.25cm, minimum width = 4cm, anchor = west] at (4.9,-2.1) {};

    \node at (5,-0.75) [draw=none, anchor=west] {\small $P_2$};
    \node at (-8.75,-0.75) [draw=none, anchor=west] {\small $P_0$, $P_1$ (global view)};
    \node at (-8.75,-4) [draw=none, anchor=west] {\small $P_0$ and $P_1$ set the shares $\sum_0^3 [\psi_i\oplus \theta_i] = [46]$};

    \node at (-8.5, -1.2) [draw=none, anchor=west] {\small $\vec{\bm{\psi}}$};
    \foreach \i in {1,...,4} {
    	\draw [-] (-8.65,-1 - 0.45*\i) -- (-7.85,-1 - 0.45*\i);  
    }  

    \node at (-8.75,-1.25 - 0.45*1) [draw=none, anchor=west] {\small $\psi_0$=$12$};
    \node at (-8.75,-1.25 - 0.45*2) [draw=none, anchor=west] {\small $\psi_1$=$46$};
    \node at (-8.75,-1.25 - 0.45*3) [draw=none, anchor=west] {\small $\psi_2$=$31$};
    \node at (-8.75,-1.25 - 0.45*4) [draw=none, anchor=west] {\small $\psi_3$=$27$};

    \foreach \i in {1,...,4} {
    	\draw [-] (-6.65,-1 - 0.45*\i) -- (-5.85,-1 - 0.45*\i);  
    } 
    \node at (-6.5, -1.2) [draw=none, anchor=west] {\small $\vec{\bm{\phi}}$};

	\node at (-6.75,-1.25 - 0.45*1) [draw=none, anchor=west] {\small $\phi_0$=31};
    \node at (-6.75,-1.25 - 0.45*2) [draw=none, anchor=west] {\small $\phi_1$=27};
    \node at (-6.75,-1.25 - 0.45*3) [draw=none, anchor=west] {\small $\phi_2$=46};
    \node at (-6.75,-1.25 - 0.45*4) [draw=none, anchor=west] {\small $\phi_3$=12};
    
    \foreach \i in {1,...,4} {
    	\draw [-] (-4.65,-1 - 0.45*\i) -- (2.25,-1 - 0.45*\i);  
    } 
    \node at (-4.75,-1.25 - 0.45*1) [draw=none, anchor=west] {\small $\phi_0$};
    \node at (-4.75,-1.25 - 0.45*2) [draw=none, anchor=west] {\small $\phi_1$};
    \node at (-4.75,-1.25 - 0.45*3) [draw=none, anchor=west] {\small $\phi_2$};
    \node at (-4.75,-1.25 - 0.45*4) [draw=none, anchor=west] {\small $\phi_3$};
	
	\draw [-] (-4.25, -1.1) -- (-4.25,-3.2); 
	\node at (-4.3,-1.25) [draw=none, anchor=west] {\small $\phi_0$}; 
    
    \foreach \i in {1,...,3} {
    	\draw [-] (-5.8 + 2*\i, -1.1) -- (-5.8 + 2*\i,-3.2); 
    	\node at (-5.85 + 2*\i,-1.25) [draw=none, anchor=west] {\small $\phi_\i$}; 
    	\node at (-5.85 + 2*\i,-1.25- 0.45*1) [draw=none, anchor=west] {\small uCMP($\psi_0$,$\psi_\i$)};
    }
    \foreach \i in {2,...,3} {
    	\node at (-5.85 + 2*\i,-1.25- 0.45*2) [draw=none, anchor=west] {\small uCMP($\psi_1$,$\psi_\i$)};
    }
    \node at (-5.85 + 2*3,-1.25- 0.45*3) [draw=none, anchor=west] {\small uCMP($\psi_2$,$\psi_3$)};

    \foreach \i in {1,...,4} {
    	\draw [-] (4.95,-1 - 0.45*\i) -- (8.85,-1 - 0.45*\i);  
    } 
    \node at (4.9,-1.25 - 0.45*1) [draw=none, anchor=west] {\small $\phi_0$};
    \node at (4.9,-1.25 - 0.45*2) [draw=none, anchor=west] {\small $\phi_1$};
    \node at (4.9,-1.25 - 0.45*3) [draw=none, anchor=west] {\small $\phi_2$};
    \node at (4.9,-1.25 - 0.45*4) [draw=none, anchor=west] {\small $\phi_3$};
	
	 \foreach \i in {0,...,3} {
    	\draw [-] (5.4+0.9*\i, -1.1) -- (5.4+0.9*\i,-3.2); 
    	\node at (5.4+0.9*\i,-1.25) [draw=none, anchor=west] {\small $\phi_\i$}; 
    }
    
    \node at (5.4+0.9*1,-1.25- 0.45*1) [draw=none, anchor=west] {\small 1};
    \node at (5.4+0.9*2,-1.25- 0.45*1) [draw=none, anchor=west] {\small \wzynote{0} };
    \node at (5.4+0.9*3,-1.25- 0.45*1) [draw=none, anchor=west] {\small 1};

    \node at (5.4+0.9*2,-1.25- 0.45*2) [draw=none, anchor=west] {\small \wzynote{0} };
    \node at (5.4+0.9*3,-1.25- 0.45*2) [draw=none, anchor=west] {\small 1};

    \node at (5.4+0.9*2,-1.25- 0.45*3) [draw=none, anchor=west] {\small \wzynote{$i^\star$} };
    \node at (5.4+0.9*3,-1.25- 0.45*3) [draw=none, anchor=west] {\small \wzynote{1} };

    \draw [->] (-7.7,-2.2) -- (-6.7,-2.2);
    \node at (-7.2,-2) [draw=none, anchor=center] {\small $\Pi(\vec{\bm{\psi}})$};

    \draw [->] (-5.7,-2.2) -- (-4.7,-2.2);
    \node at (-5.2,-2) [draw=none, anchor=center] {\small $\text{uCMP}$};

    \node at (3.7,-0.75) [draw=none, anchor=center] {\small $\{\text{uCMP}_{i,j}\}$};
    \draw [->] (2.6,-1) -- (4.8,-1);

    \draw [->] (4.8,-3) -- (2.6,-3);
    \node at (3.7,-2.75) [draw=none, anchor=center] {\small $[\vec{\bm{\theta}}]$};

    \node at (4.75,-3.4) [draw=none, anchor=west] {\tiny $\forall i<i^\star$, $\text{uCMP}_{i,i^\star}$ is 0/NULL.};
    \node at (4.75,-3.6) [draw=none, anchor=west] {\tiny $\forall j>i^\star$, $\text{uCMP}_{i^\star,j}$ is 1/NULL.};
    \node at (4.75,-3.8) [draw=none, anchor=west] {\tiny $i^\star=2$, $\vec{\bm{\theta}}=\{0,0,1,0\}$};

\end{tikzpicture}
\caption{MAX example.}
\label{fig:max}
\end{figure*}
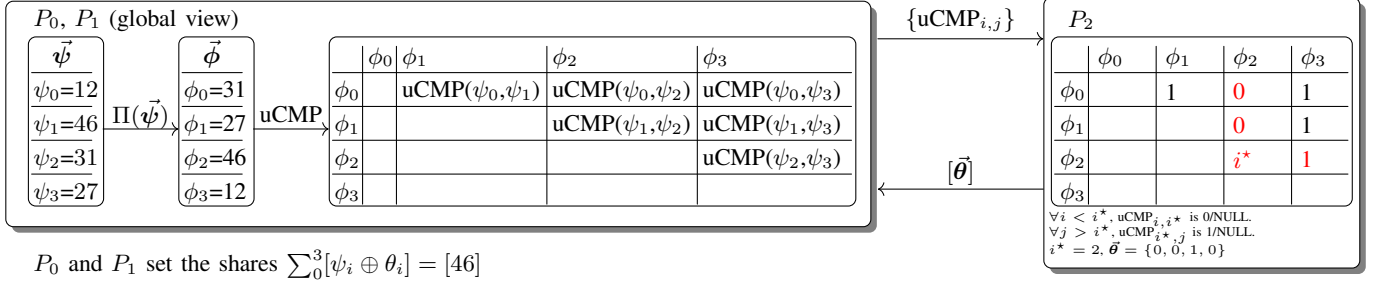
\begin{algorithm}[t]
	\hspace*{\algorithmicindent} \textbf{Input}: the shares of $\vec{\bm{\psi}}=(\psi_1,\cdots,\psi_n)$\\
	\hspace*{\algorithmicindent} \textbf{Output}: the shares of $\text{MAX}(\vec{\bm{\psi}})$ 
    \begin{algorithmic}[1]
    \State $P_0$ and $P_1$ shuffle the input array $\vec{\bm{\psi}}$ using the same shuffle-seed generated from $\textsf{seed}_{01}$ and get $\vec{\bm{\phi}}$, then compare each pair of elements in $\vec{\bm{\phi}}$ using uCMP, without duplication. 
    \State $P_0,P_2$ and $P_1,P_2$ generate the shares of $n$ triples from $\textsf{seed}_{02}$ and $\textsf{seed}_{12}$, respectively.
    \State $P_0$ sends $[d_i]_0: = [\phi_i]_0 - [a_i]_0$ to $P_1$, and $P_1$ sends $[d_i]_1: = [\phi_i]_1 - [a_i]_1$ to $P_0$, $\forall i\in[1,n]$.
    \State $P_0$ and $P_1$ send the requests of $\text{uCMP}(\phi_i,\phi_j)$ ($\forall i,j\in[1,n]$ and $i \neq j$) to $P_2$.
    \State $P_2$ determines the target indexes $i^\star$ satisfying $\forall i<i^\star$, $\text{uCMP}(\phi_i,\phi_{i^*})$ is 0 or NULL, and $\forall j>i^\star$, $\text{uCMP}(\phi_{i^*},\phi_j)$ is 1 or NULL. \label{alg:max:step:relu} 
    \State $P_2$ constructs a weight-one vector $\vec{\bm{\theta}}$, in which $\theta_{i^\star}=1$ and $\theta_{i}=0,\forall i\neq i^\star,i\in[1,n]$. 
    \State $P_2$ sets $e_i:=\theta_i - b_i$, and sends $e_i$ to $P_0$, $e_i$ and $[c_i]_1$ to $P_1$, $\forall i\in[1,n]$.
	\State $P_0$ and $P_1$ set $d_i:=[d_i]_0 + [d_i]_1$, $\forall i\in[1,n]$, and compute the shares $[\text{MAX}(\vec{\bm{\psi}})]=\sum_{i=1}^n [\phi_i\cdot \theta_i] = \sum_{i=1}^n(d_i\cdot e_i+d_i[b_i]+e_i[a_i]+[c_i])$
    \end{algorithmic}
  \caption{MAX protocol.}
  \label{alg:max}
\end{algorithm}

\subsection{MAX}
In this section, we present two types of Maxpool protocols, namely, binary-tree-based Maxpool~\footnote{The binary-tree-based Maxpool simply invokes $\mathcal{O}(n)$ times of MAX2 (Eq.~\ref{eq:max}) for $O(log_2 n)$ rounds.} and 2-round Maxpool, each with its own advantages. While the bandwidth is not saturated and the latency is dominating the performance, e.g. under LAN settings with a small batch size for small $n$, the 2-round Maxpool protocol provides better performance. In contrast, the binary-tree-based Maxpool protocol performs better while the network bandwidth is not rich, e.g. under WAN settings or LAN settings with a large batch size. The communication complexity of the 2-round and binary-tree-based maxpool protocols are $\mathcal{O}(n^2)$ and $\mathcal{O}(n)$ respectively.

The 2-round maxpool protocol invokes the unblind version of the CMP protocol, in which $P_2$ learns the exact relation between the inputs. We denote the unblind CMP protocol as $\text{uCMP}(x,y):=\text{uDReLU}(x-y)$, which simply omits the step~\ref{alg:drelu:step:reverse} and~\ref{alg:drelu:step:final}  of Alg.~\ref{alg:drelu}, i.e., without randomly negating the input and output, while keeping the remaining steps identical.

Based on the uCMP protocol, we can let $P_2$ compare a list of randomly shuffled inputs which leads to the functionality of determining the maximum element. In particular, $P_0$ and $P_1$ first randomly shuffle the $n$-length ($n>2$) input vector $\vec{\bm{\psi}}$ to get $\vec{\bm{\phi}}$ using $\textsf{seed}_{01}$. Then, $P_0$ and $P_1$ compare each pair of items in $\vec{\bm{\phi}}$ by invoking the uCMP protocol for $\frac{n \cdot (n-1)}{2}$ times. These invocations form an upper triangular matrix as illustrated in Fig.~\ref{fig:max}. After computing the output of the comparison matrix $\{\text{uCMP}_{i,j}\}$, $P_2$ looks for index $i^*$ such that $\forall i<i^\star$, $\text{uCMP}_{i,i^*} = 0$ or NULL and $\forall j>i^\star$, $\text{uCMP}_{i^*,j} = 1$ or NULL. $P_2$ then constructs a result weight-one indicator vector $\vec{\bm{\theta}}$ such that $\theta_{i^\star}=1$, and $\theta_{i}=0,\forall i\neq i^\star,i\in[1,n]$. Then, $P_2$ sends the shares of $\vec{\bm{\theta}}$ to $P_0$ and $P_1$. $P_0$ and $P_1$ finally output the shares of the maximum element, i.e., the shares of $\text{MAX}(\vec{\bm{\psi}}) := \sum_{i=1}^n \phi_i\cdot \theta_i$. Notice that the round-collapsing optimization of ReLU could also be applied in this case. We present the details of the algorithm in Alg.~\ref{alg:max}.

\subsection{MIN, SORT, and MED Protocols} \label{sub:minetc}
Building on the MAX protocol, we could design protocols for the minimum, sorting, and median functionalities. The MIN protocol is similar to the MAX protocol where we only need to change $P_2$'s determination strategy (step~\ref{alg:max:step:relu} of Alg.~\ref{alg:max}). To determine the minimum element, $P_2$ looks for the index $i^\star$ subject to $\forall i<i^\star$, $\text{uCMP}_{i,i^\star} = 1$ and $\forall j>i^\star$, $\text{uCMP}_{i^\star,j} = 0$.

\begin{figure}[h]
  \centering
  \begin{tikzpicture}[x=1cm,y=1cm,cap=round,align=center,
    fact/.style={fill=red!20, minimum size=5mm}]
    \matrix (m1) [nodes={draw,minimum size=5mm}] at (-3,0)
    {
    \node { }; & \node{1}; & \node[fact] {0}; & \node {1}; \\
    \node { }; & \node{ }; & \node[fact] {0}; & \node {1};\\
    \node[fact] { }; & \node[fact] { }; & \node[fact] { }; & \node[fact] {1};\\
    \node { }; & \node{ }; & \node[fact] { }; & \node { };\\
    };

    \matrix (m2) [nodes={draw,minimum size=5mm}] at (0,0)
    {
    \node[fact] { }; & \node[fact] {1}; & \node[fact] {1}; \\
    \node[fact] { }; & \node{ }; & \node {1};\\
    \node[fact] { }; & \node{ }; & \node { };\\
    };

    \draw [->] (m1) -- (m2);

    \matrix (m3) [nodes={draw,minimum size=5mm}] at (3,0)
    {
    \node[fact] { }; & \node[fact] {1}; \\
    \node[fact] { }; & \node{ }; \\
    };

    \draw [->] (m2) -- (m3);

\end{tikzpicture}
\caption{$P_2$ repeatedly looks for the index of the maximum value and removes the corresponding row and column in the descending order SORT protocol.}
  \label{fig:sortmatrix}
\end{figure}
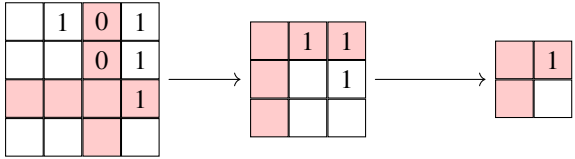

For the (descending) SORT protocol, $P_2$ should continuously look for the maximum element in the matrix. After determining the maximum element (corresponding the indexes $i^\star$), $P_2$ removes the $i^\star$-th row and the $i^\star$-th column from the matrix. Then, it repeats the above process until the matrix is empty. Fig.~\ref{fig:sortmatrix} illustrates this process (the row and column to be removed are marked in red). 

After repeating $n-1$ times of the above process, $P_2$ could construct a permutation matrix $\bm{M}$, in which $m_{i,j}=1$ if $\psi_i$ is the $j$-th maximum value and $m_{i,j}=0$ otherwise. Finally, $P_0$ and $P_1$ multiply $\bm{M}$ by $\vec{\bm{\phi}}$ in the secret sharing form to acquire the shared sorted arrays. 

Since the MAX protocol involves multiplications of the $i$-th row of the matrix and $\vec{\bm{\phi}}$ for $\forall i\in[1,n]$, we only need $n$ numbers of triples $b$ to mask $\vec{\bm{\phi}}$. Also notice that we can apply the balancing technique proposed in Sect.~\ref{sec:preliminary} to reduce the amount of one-way communication.

Similarly, in the ascending order SORT protocol, $P_2$ continuously searches for the minimum element. Note that, $P_2$ could simultaneously search for the maximum and minimum element in both the descending and ascending order SORT protocols, then eliminate two rows and two columns at the same time, which accelerates the sorting algorithm.

For the MED protocol, we could let $P_2$ repeat the step~\ref{alg:max:step:relu} in Alg.~\ref{alg:max} for $\lceil \frac{n}{2} \rceil$ times. Then, $P_2$ constructs the indicator vector to reflect the index corresponding to the median element. The remaining steps of MED are the same as those of the MAX protocol.

\begin{table*}[t!]
  \caption{The communication overhead of our protocols. (DReLU includes DReLU/MSB/BitExt/CMP. ReLU includes ReLU/ABS/Dynamic ReLU/MAX2/MIN2/Funnel ReLU. $\ell$: the bit length of an integer ring $\mathbb{Z}_q=[0,q-1]$ ($\ell:=\log_2q$). $\ell_x$: the precision of the input. $m+2$ piecewise for PLU. $n$ inputs for MAX/MIN/SORT/MED. Prep.: preprocessing. Rnd.: round. $\text{Comm}_{ij}$: the communication overhead from $p_i$ to $p_j$, counting in bit. $\textbf{Comm}_{\text{total}}$: the total communication overhead among the three participants. $\text{Comm}_{02}$ or $\text{Comm}_{12}$ consumes the largest amount of communication in our protocols, which is the dominant one-way communication.)}
  \label{tab:comm}
  \centering
    \linespread{1.25}
    \footnotesize
      \begin{tabular*}{17.75cm}{@{\extracolsep{\fill}} p{1cm} p{0.5cm} p{0.5cm} p{1cm} p{2cm} p{2cm} p{1cm} p{1.3cm} p{1.3cm} l }
        \toprule
        \textbf{Protocol} & \textbf{Prep.} & \textbf{Rnd.} & $\textbf{Comm}_{01}$ & $\textbf{Comm}_{02}$ & $\textbf{Comm}_{12}$ & $\textbf{Comm}_{10}$ & $\textbf{Comm}_{20}$ & $\textbf{Comm}_{21}$ & $\textbf{Comm}_{\text{total}}$ \\
        \hline
        DReLU & No & 2 & N.A. & $(\ell_x+2)\ell$ & $(\ell_x+2)\ell$ & N.A. & $\ell$ & $\ell$ & $(2\ell_x+4)\ell$\\
        ReLU & No & 2 & $\ell$ & $(\ell_x+2)\ell$ & $(\ell_x+2)\ell$ & $\ell$ & $\ell$ & $2\ell$ & $(2\ell_x+9)\ell$\\
	      PLU   & No & 2 & $\ell$ & $(m+1)(\ell_x+2)\ell$  & $(m+1)(\ell_x+2)\ell$ & $\ell$ & $\lceil \frac{5m+10}{2} \rceil\ell$ & $\lfloor \frac{5m+10}{2} \rfloor\ell$ & $((2m+1)\ell_x+9m +16)\ell$ \\
        RELU6 & No & 2 & $\ell$ & $2(\ell_x+2)\ell$  & $2(\ell_x+2)\ell$ & $\ell$ & $\lceil \frac{15}{2} \rceil\ell$ & $\lfloor \frac{15}{2} \rfloor\ell$ & $(4\ell_x+21)\ell$ \\
        MAX/MIN & No & 2 & $n\ell$ & $\frac{n(n-1)}{2}(\ell_x+2)\ell$  & $\frac{n(n-1)}{2}(\ell_x+2)\ell$ & $n\ell$ & $\lceil \frac{3n}{2} \rceil\ell$ & $\lfloor \frac{3n}{2} \rfloor\ell$ & $((2+\ell_x)n^2 + (3-\ell_x)n)\ell$ \\
        SORT & No & 2 & $n\ell$ & $\frac{n(n-1)}{2}(\ell_x+2)\ell$  & $\frac{n(n-1)}{2}(\ell_x+2)\ell$ & $n\ell$ & $\lceil \frac{3n^2}{2} \rceil\ell$ & $\lfloor \frac{3n^2}{2} \rfloor\ell$ & $((5+\ell_x)n^2 -\ell_xn)\ell$ \\
        MED & No & 2 & $n\ell$ & $\frac{n(n-1)}{2}(\ell_x+2)\ell$  & $\frac{n(n-1)}{2}(\ell_x+2)\ell$ & $n\ell$ & $\lceil \frac{3n}{2} \rceil\ell$ & $\lfloor \frac{3n}{2} \rfloor\ell$ & $((2+\ell_x)n^2 + (3-\ell_x)n)\ell$ \\
    \bottomrule
    \end{tabular*}
\end{table*}

\section{Evaluation and Experiments} \label{sec:evaluate}
We first evaluate the communication overhead of our proposed protocols in Sect.~\ref{sub:comoverhead}, and then run experiments to show the practical performance of our protocols in Sect.~\ref{sub:experiment}.

\subsection{Evaluation} \label{sub:comoverhead}
We calculate the theoretical communication overhead in Tab.~\ref{tab:comm} to have a better view of how the communication overhead dominates the performance of an MPC protocol. $\text{Comm}_{01}$, $\text{Comm}_{02}$, and $\text{Comm}_{12}$ denote the uni-direction communication overhead between $P_0$ and $P_1$, $P_0$ and $P_2$, and $P_0$ and $P_1$, respectively, while $\text{Comm}_{10}$, $\text{Comm}_{20}$, and $\text{Comm}_{21}$ correspond to the reverse directions. $\text{Comm}_{total}$ donates the total communication overhead for each protocol. If more than one triple is used in a protocol, we consider a setting in which $P_2$ distributes around half of $[c]$'s to $P_0$, while delivering the other half to $P_1$. Among our protocols, $\text{Comm}_{02}$ or $\text{Comm}_{12}$ has the greatest communication overhead comparing with other one-way communication pass, i.e., $\text{Comm}_{10}$, $\text{Comm}_{01}$, etc. Hence, for fairness we use $\text{Comm}_{02}$ to compare with that of other related protocols in Tab.~\ref{tab:compare}.

\subsection{Experiments} \label{sub:experiment}
All experiments are carried out under LAN settings, i.e., three cloud virtual machines (VM) in the same region, and we select $q = 2^{64}$ and $\ell_x = 13$ as our system settings. Each VM has 8 CPUs with 2.2GHz (Intel(R) Xeon(R) Gold 6161) and 8GiB memory. The 64Byte-size PING latency between each two VMs is around 250$\mu$s to 500$\mu$s. The bandwidth (measured by $\textsf{netperf}$) between each participant ranges from 1250-1780MBytes/s.

We run experiments on our protocols and two state-of-the-art protocols, Falcon~\cite{WTB+-21,falcon-code} and Edabits~\cite{DSM+-20,MP-SPDZ}. The results are shown in Tab.~\ref{tab:performance1} and~\ref{tab:performance2}.~\footnote{We select the honest-majority and passive-secure settings in Edabits and Falcon. Edabits~\cite{DSM+-20} does not have a Maxpool design.} We also carry out experiments on EQ/ReLU6/MED4/SORT4 and the results are shown in Tab.~\ref{tab:performance3}. We provide comparison between our DReLU/ReLU/MAX4/MAX9 and the aforementioned works, and a brief analysis in the following paragraphs.

\noindent \textbf{DReLU and ReLU.} When the batch size is 100,000, we achieve more than 390,000 DReLU and 370,000 ReLU operations per second, respectively. It is worth mentioning that the source codes of Falcon~\cite{falcon-code} and Edabits~\cite{MP-SPDZ} do not contain the preprocessing which is heavily computed, and hence, the experiment results of the latency also do not include the preprocessing. By comparing the latency of our protocols (around 340 ms) with the latency for online phase only of Falcon~\cite{falcon-code} and Edabits~\cite{MP-SPDZ}, our work has a one or two orders of magnitude improvement, respectively, without batching.

\noindent \textbf{Maxpool.} Our Maxpool protocols achieve the throughput rate of 110,000 or 41,000 operations per second, for binary-tree-based MAX4 or MAX9, respectively. When the bandwidth is not saturated, i.e., small batch size and small $n$, both our 2-round and binary-tree-based Maxpool protocols outperform Falcon's protocols. When the batch size is large, our binary-tree-based Maxpool protocols still have better performance than Falcon's. Once again, the latency recorded from the Falcon protocols does not contain its heavy preprocessing.

\noindent \textbf{Estimated E2E inference.} Based on the above experiment results, we could give an estimation of the end-to-end improvement in PPML by relating the obtained speed-ups on ReLU back with Fig.~\ref{fig:cryptgpu}. We use Cifar10\_vgg16\_32 as an example, such a model contains 16 ReLU layers each with batch size ranging from 32$\times$100 to 32$\times$65,535. Using our experiment results, we calculate the total time needed for operating all ReLU layers using our protocols is around 15.3 s. By substituting this time for the time used in Figure 1, i.e., 106 s (time for all ReLU operations) out of 126 s (time for the whole inference), we could achieve a $3.6\times$ of improvement on the complete interface. However, this estimation omits the difference between CPU and GPU platforms. We are working on the GPU integration of Bicoptor protocols and will provide a more rigorous experiment result on the end-to-end improvement in future work.

\section{Conclusion} \label{sec:conclude}
We propose Bicoptor, which includes several two-round three-party computation protocols without preprocessing for PPML. We innovate a novel sign determination protocol, which relies on a clever use of the truncation protocol. Bicoptor supports the DReLU, Equality, ABS, ReLU, Dynamic ReLU (Leaky ReLU, PReLU, RReLU), ReLU6, Piecewise Linear Unit (PLU), MAX, MIN, SORT, and median (MED) functions. The experiments exhibit our fast performance, which has a one or two orders of magnitude improvement to ReLU in the state-of-the-art works, i.e., Falcon or Edabits, respectively. 

\section*{Acknowledgment}
We wish to express our gratitude to Xu et al.~\cite{XLH24} for their detail analysis and research on our protocol, along with the security concerns he raised. We would like to thank Zhongkai Li, Wenyuan Tian, Xianggui Wang, and Hao Guo for their great help in the implementations of Bicoptor. Yu Yu was supported by the National Natural Science Foundation of China (Grant Nos.62125204, 92270201 and 61872236). Yu Yu also acknowledges the support from the XPLORER PRIZE.

\begin{table*}[t!]
  \caption{The performance of Bicoptor, Falcon~\cite{WTB+-21} and Edabits~\cite{DSM+-20} DReLU and ReLU protocols. (Lat.: latency. ALat.: amortized latency. Thru.: throughput rate. $q=2^{64},\ \ell_x = 13$.)}
  \label{tab:performance1}
  \centering
    \linespread{1.25}
    \footnotesize
    \begin{tabular*}{17.75cm}{@{\extracolsep{\fill}} p{0.5cm} l p{1.5cm} p{1.5cm} p{1.5cm} p{1.5cm} p{1.5cm} p{1.5cm} p{1.5cm}}
        \toprule
        \textbf{Batch} & \multirow{2}*{\textbf{Protocol}} & \multicolumn{3}{c}{\textbf{DReLU}} & \multicolumn{3}{c}{\textbf{ReLU}} \\
        \cline{3-8}
        \textbf{Size} & & Lat.($\mu$s) & ALat.($\mu$s) & Thru.(ops/s) & Lat.(us) & ALat.(us) & Thru.(ops/s) \\
        \hline
        \multirow{3}*{1} &
        Bicoptor & 330.2  & 330.2  & 3028.6  & 340.0  & 340.0  & 2941.1                     \\
        & Falcon~\cite{WTB+-21} & 1005.8  & 1005.8  & 994.3  & 1184.4  & 1184.4  & 844.3    \\
        & Edabits~\cite{DSM+-20} & 22104.7  & 22104.7  & 45.2  & 24137.9  & 24137.9  & 41.4 \\
        \hline
        \multirow{3}*{$10^2$} &
        Bicoptor & 703.0  & 7.0  & 142243.3  & 705.4  & 7.1  & 141760.4                     \\ 
        & Falcon~\cite{WTB+-21} & 1220.0  & 12.2  & 81969.5  & 1374.6  & 13.7  & 72748.4    \\ 
        & Edabits~\cite{DSM+-20} & 179186.0  & 1791.9  & 558.1  & 180822.0  & 1808.2 & 553.0\\
        \hline
        \multirow{3}*{$10^3$} &
        Bicoptor & 3078.1  & 3.1  & 324872.2  & 3428.9  & 3.4  & 291635.4                   \\
        & Falcon~\cite{WTB+-21} & 3515.7  & 3.5  & 284435.7  & 4118.5  & 4.1  & 242806.8    \\ 
        & Edabits~\cite{DSM+-20} & 486465.7  & 486.5  & 2055.6  & 566871.0  & 566.9 & 1764.1\\
        \hline
        \multirow{3}*{$10^4$} &
        Bicoptor & 25969.9  & 2.6  & 385061.2  & 31531.4  & 3.2  & 317144.5  \\ 
        & Falcon~\cite{WTB+-21} & 22266.5  & 2.2  & 449105.2  & 33749.6  & 3.4  & 296299.8  \\
        & Edabits~\cite{DSM+-20} & 744525.3  & 74.5  & 13431.4  & 795877.7  & 79.6  & 12564.7 \\
        \hline
        \multirow{3}*{$10^5$} &
        Bicoptor & 253397.3  & 2.5  & 394637.1  & 267570.0  & 2.7  & 373734.0                \\
        & Falcon~\cite{WTB+-21} & 322137.0  & 3.2  & 310426.9  & 333221.0  & 3.3  & 300101.1\\
        & Edabits~\cite{DSM+-20}& 2368510.0  & 23.7  & 42220.6  & 2563846.7  & 25.6  & 39003.9  \\
    \bottomrule
    \end{tabular*}
\end{table*}
\begin{table*}[t!]
  \caption{The performance of binary-tree and 2-round Bicoptor MAX4/MAX9, and Falcon~\cite{WTB+-21} MAX4/MAX9 protocols. (Lat.: latency. ALat.: amortized latency. Thru.: throughput rate. $q=2^{64},\ \ell_x = 13$. Bicoptor-Binary: the binary-tree-based maxpool Bicoptor protocols.)}
  \label{tab:performance2}
  \centering
    \linespread{1.25}
    \footnotesize
      \begin{tabular*}{17.75cm}{@{\extracolsep{\fill}} p{0.5cm} l p{1.5cm} p{1.5cm} p{1.5cm} p{1.5cm} p{1.5cm} p{1.5cm} p{1.5cm}}
        \toprule
        \textbf{Batch} & \multirow{2}*{\textbf{Protocol}} & \multicolumn{3}{c}{\textbf{MAX4}} & \multicolumn{3}{c}{\textbf{MAX9}} \\
        \cline{3-8}
        \textbf{Size} & & Lat.($\mu$s) & ALat.($\mu$s) & Thru.(ops/s) & Lat.(us) & ALat.(us) & Thru.(ops/s) \\
        \hline
        \multirow{3}*{1} &
        Bicoptor-Binary & 651.8  & 651.8  & 1534.2  & 1617.0  & 1617.0  & 618.4  \\
        & Bicoptor-2-round & 341.6  & 341.6  & 2927.3  & 455.1  & 455.1  & 2197.5  \\
        & Falcon~\cite{WTB+-21}& 3703.7  & 3703.7  & 270.0  & 10449.5  & 10449.5  & 95.7 \\
        \hline
        \multirow{3}*{$10^2$} &
        Bicoptor-Binary & 1718.1  & 17.2  & 58202.3  & 4211.1  & 42.1  & 23747.0  \\
        & Bicoptor-2-round & 2179.3  & 21.8  & 45886.4  & 9673.2  & 96.7  & 10337.8  \\
        & Falcon~\cite{WTB+-21}& 4509.4  & 45.1  & 22176.1  & 14175.3  & 141.8  & 7054.5 \\
        \hline
        \multirow{3}*{$10^3$} &
        Bicoptor-Binary & 9783.3  & 9.8  & 102214.6  & 26420.8  & 26.4  & 37849.0 \\
        & Bicoptor-2-round & 16399.8  & 16.4  & 60976.5  & 90154.9  & 90.2  & 11092.0 \\
        & Falcon~\cite{WTB+-21}& 12635.9  & 12.6  & 79139.4  & 34885.1  & 34.9  & 28665.6  \\
        \hline
        \multirow{3}*{$10^4$} &
        Bicoptor-Binary &  87470.8  & 8.7  & 114323.8  & 244932.0  & 24.5  & 40827.7 \\
        & Bicoptor-2-round & 165940.2  & 16.6  & 60262.7  & 1015628.1  & 101.6  & 9846.1 \\
        & Falcon~\cite{WTB+-21}& 111099.7  & 11.1  & 90009.3  & 297128.0  & 29.7  & 33655.5 \\
        \hline
        \multirow{3}*{$10^5$} &
        Bicoptor-Binary & 858400.8  & 8.6  & 116495.7  & 2404498.0  & 24.0  & 41588.7 \\
        & Bicoptor-2-round & 1723884.6  & 17.2  & 58008.5  & 10410762.7  & 104.1  & 9605.4 \\
        & Falcon~\cite{WTB+-21}& 1063411.3  & 10.6  & 94037.0  & 2802530.0  & 28.0  & 35682.0 \\
    \bottomrule
    \end{tabular*}
\end{table*}
\begin{table*}[t!]
  \caption{The performance of Bicoptor EQ, ReLU6, MED4, and SORT6 protocols. (Lat.: latency. ALat.: amortized latency. Thru.: throughput rate. $q=2^{64},\ \ell_x = 13$.)}
  \label{tab:performance3}
  \centering
    \linespread{1.25}
    \footnotesize
      \begin{tabular*}{17.75cm}{@{\extracolsep{\fill}} p{0.5cm} p{1.5cm} p{1.5cm} p{1.5cm} p{1.5cm} p{1.5cm} p{1.5cm} p{1.5cm}}
        \toprule  
        \textbf{Batch} & \multicolumn{3}{c}{\textbf{EQ}} & \multicolumn{3}{c}{\textbf{ReLU6}} \\
        \cline{2-7}
        \textbf{Size} & Lat.($\mu$s) & ALat.($\mu$s) & Thru.(ops/s) & Lat.(us) & ALat.(us) & Thru.(ops/s) \\
        \hline
        1      & 334.4  & 334.4  & 2990.1  & 355.0  & 355.0  & 2817.1 \\
        $10^2$ & 936.8  & 9.4  & 106743.3  & 1104.9  & 11.0  & 90505.5\\
        $10^3$ & 5931.9  & 5.9  & 168578.9  & 6685.8  & 6.7  & 149571.5\\
        $10^4$ & 63747.3  & 6.4  & 156869.3  & 62503.6  & 6.3  & 159990.8 \\
        $10^5$ & 528272.7  & 5.3  & 189296.2  & 620112.0  & 6.2  & 161261.2 \\
        \hline
        \textbf{Batch} & \multicolumn{3}{c}{\textbf{MED4}} & \multicolumn{3}{c}{\textbf{SORT4}} \\
        \cline{2-7}
        \textbf{Size} & Lat.($\mu$s) & ALat.($\mu$s) & Thru.(ops/s) & Lat.(us) & ALat.(us) & Thru.(ops/s) \\
        \hline
        1      & 361.9  & 361.9  & 2763.3  & 381.6  & 381.6  & 2620.5  \\
        $10^2$ & 2282.3  & 22.8  & 43815.8  & 2506.5  & 25.1  & 39895.7 \\
        $10^3$ & 17488.1  & 17.5  & 57181.7  & 21285.4  & 21.3  & 46980.6 \\
        $10^4$ & 169422.3  & 16.9  & 59024.1  & 193220.6  & 19.3  & 51754.3\\
        $10^5$ & 1906699.0  & 19.1  & 52446.7  & 2425505.4  & 24.3  & 41228.5 \\
    \bottomrule
    \end{tabular*}
\end{table*}

\bibliographystyle{IEEEtran}
\bibliography{IEEEabrv,main}

~
~
\newpage
\appendices
\section{Decimal Arithmetic with Fixed-point Computation} \label{app:fprn}
In a practical PPML application, the inputs and coefficients could be decimal numbers. To express a decimal number, we consider a typical case in which the bit length of an integer is the same as that of the fractional part, i.e., the last $\frac{\ell_x}{2}$ bits are used to represent the fractional part. Hence, $x:=\sum_{i=0}^{\ell_x-1} x_i\cdot 2^{i - \frac{\ell_x}{2}}$ where $x_i$ is the $i$-th bit of $x$. This expression works for both positive and negative decimal numbers. After a multiplication, a truncation is required to recover the original precision. Tab.~\ref{tab:fprn} shows an example.
\begin{table}[t!]
  \caption{An example of the fixed-point computation for a decimal multiplication ($q=2^{40},\ \ell_x = 16$).}
  \label{tab:fprn}
  \centering
    \linespread{1.25}
    \footnotesize
      \begin{tabular*}{8.5cm}{@{\extracolsep{\fill}} l p{7.5cm} }
        \toprule
        \textbf{Decimal} & \textbf{Binary} \\
        \hline
        $\alpha$=10.82421875                           & 00000000\_00000000\_00000000\_00001010\_11010011 \\
        $\beta$=6.2265625                              & 00000000\_00000000\_00000000\_00000110\_00111010 \\
        $\gamma$=$-$6.2265625                           & 11111111\_11111111\_11111111\_11111011\_11000110 \\
        TRC($\alpha\cdot\beta$,8)  &  \multirow{2}{7.5cm}{00000000\_00000000\_00000000\_01000011\_01100101}\\
        =67.39453125 & \\
        TRC($\alpha\cdot\gamma$,8)  & \multirow{2}{7.5cm}{11111111\_11111111\_11111111\_11101110\_00100010}\\
        =$-$67.39453125 & \\
        \bottomrule
      \end{tabular*}
\end{table}

\section{Non-linear Function Definition in PPML} \label{app:functions}
\noindent \textbf{Bit-wise univariate function}. From DReLU, the bit extraction (BitExt) and the most significant bit (MSB) functions are obtained as Eq.~\ref{eq:bitextandmsb}.
\begin{equation} \label{eq:bitextandmsb}
\begin{aligned}
&\text{BitExt}(x) = \text{MSB}(x) = \left\{
\begin{aligned}
	&0, \quad x\geq0 \\
	&1, \quad x<0 \\
\end{aligned}
\right . \\
=& 1 - \text{DReLU}(x) 
\end{aligned}
\end{equation}

\noindent \textbf{Univariate function with two segments}. Other functions could also be derived from DReLU, such as ABS (Eq.~\ref{eq:abs}), ReLU (Eq.~\ref{eq:relu}), and Dynamic ReLU (Eq.~\ref{eq:dynamicrelu}).~\footnote{Leaky ReLU, PReLU, and RReLU are the special cases of Dynamic ReLU. For Leaky ReLU, $\alpha_0=0.001,\alpha_1=1$. For PReLU, $\alpha_0$ is a pre-trained constant and $\alpha_1=1$. For RReLU, $\alpha_0$ is a random constant and $\alpha_1=1$.}
\begin{equation} \label{eq:abs}
\begin{aligned}
&\text{ABS}(x) = \left\{ 
\begin{aligned}
	& x, \quad x\geq0 \\
	& -x, \quad x<0 \\
\end{aligned}
\right . \\
=& \text{DReLU}(x)\cdot x + (\text{DReLU}(x)-1)\cdot x\\
=& (2\cdot\text{DReLU}(x) - 1)\cdot x
\end{aligned}
\end{equation}

\begin{equation} \label{eq:dynamicrelu}
\begin{aligned}
& \text{Dynamic ReLU}(x) = \left\{ 
\begin{aligned}
	& \alpha_1 \cdot x  \quad x\geq0 \quad \\
	& \alpha_0 \cdot x  \quad x<0 \quad \\
\end{aligned}
\right . \\
&(\alpha_0, \alpha_1 \text{ are two constants.}) \\
=& \alpha_1\cdot \text{DReLU}(x)\cdot x + \alpha_0\cdot (1-\text{DReLU}(x))\cdot x\\
=& (\alpha_0 + (\alpha_1-\alpha_0)\cdot\text{DReLU}(x))\cdot x\\
\end{aligned}
\end{equation}

\noindent \textbf{Comparison function}. From DReLU, we can also formulate the comparison function(CMP) and the Equality function (Eq.~\ref{eq:equality}).
\begin{equation} \label{eq:cmp}
\text{CMP}(x_0,x_1) = \left\{
\begin{aligned}
	&1 \quad x_0\geq x_1 \\
	&0 \quad x_0<x_1 \\
\end{aligned}
\right . = \text{DReLU}(x_0-x_1)
\end{equation}

\begin{equation} \label{eq:equality}
\begin{aligned}
&\text{Equality}(x_0,x_1) = \left\{
\begin{aligned}
	&1 \quad x_0 = x_1 \\
	&0 \quad x_0\neq x_1 \\
\end{aligned}
\right . \\
=& 1 - \text{CMP}(x_0,x_1)\oplus\text{CMP}(x_1,x_0) \\
=& 1 - \text{DReLU}(x_0-x_1)\oplus\text{DReLU}(x_1-x_0)
\end{aligned}
\end{equation}

The piecewise linear unit (PLU, Eq.~\ref{eq:plu}) function has more than two input intervals, utilizing the CMP function. 

\begin{equation} \label{eq:plu}
\begin{aligned}
& \text{PLU}(x) = \left\{ 
\begin{aligned}
	& \alpha_{m+1}\cdot x + \beta_{m+1}  ,\quad \gamma_m<=x \quad \\
  & \alpha_m\cdot x + \beta_m  ,\quad \gamma_{m-1}<=x<\gamma_m \quad \\
	& ... \\
	& \alpha_j\cdot x + \beta_j  ,\quad \gamma_{j-1}<=x<\gamma_j \quad \\
	& ... \\
	& \alpha_1\cdot x + \beta_1  ,\quad \gamma_0 <=x < \gamma_1 \quad \\
	& \alpha_0\cdot x + \beta_0  ,\quad x < \gamma_0 \quad 
\end{aligned}
\right . \\
&(m+2\ \text{piecewise.}\ \alpha_j\ \text{and}\ \beta_j\ (\forall j\in[0,m+1]),\\
&\text{and}\ \gamma_j\ (\forall j\in[0,m])\ \text{are constants}.)\\
=& (\text{CMP}(x, \gamma_m)\oplus 0)\cdot (\alpha_{m+1}\cdot x + \beta_{m+1}) + ...\\
&+ (\text{CMP}(x, \gamma_{j-1}) \oplus \text{CMP}(x, \gamma_j)) \cdot (\alpha_j\cdot x + \beta_j) \\
&+ ... + (1 \oplus \text{CMP}(x, \gamma_0))\cdot (\alpha_0\cdot x + \beta_0) \\
=& (\text{DReLu}(x-\gamma_m)\oplus 0)\cdot (\alpha_{m+1}\cdot x + \beta_{m+1}) + ...\\
&+ (\text{DReLu}(x-\gamma_{j-1}) \oplus \text{DReLu}(x-\gamma_j))\cdot (\alpha_j\cdot x + \beta_j)\\
&+ ... +(1 \oplus \text{DReLu}(x-\gamma_0))\cdot (\alpha_0\cdot x + \beta_0)
\end{aligned}
\end{equation}

The ReLU6 (Eq.~\ref{eq:relu6}) function is a special case of PLU with three input intervals, i.e., $m=1,\alpha_0=\beta_0=\beta_1=\alpha_2=\gamma_0=0,\alpha_1=1,\beta_2=\gamma_2=6$.
\begin{equation} \label{eq:relu6}
\begin{aligned}
& \text{ReLU6}(x) = \left\{ 
\begin{aligned}
	& 6  \quad 6<=x \quad \\
	& x  \quad 0<=x<6 \quad \\
	& 0  \quad x<0 \quad \\
\end{aligned}
\right . \\
=& (\text{DReLu}(x-6)\oplus 0)\cdot 6  \\
&+ (\text{DReLu}(x-0) \oplus \text{DReLu}(x-6))\cdot x \\
&+ (1 \oplus \text{DReLu}(x))\cdot 0
\end{aligned}
\end{equation}

The CMP can also derive the maximum (MAX2, Eq.~\ref{eq:max}) and minimum (MIN2, Eq.~\ref{eq:min}) functions for two inputs. 

\begin{equation} \label{eq:max}
\begin{aligned}
& \text{MAX2}(x,y) = \left\{ 
\begin{aligned}
	& x  \quad x\geq y \quad \\
	& y  \quad x<y \quad \\
\end{aligned}
\right . \\
=& \text{CMP}(x,y)\cdot x + (1-\text{CMP}(x,y))\cdot y\\
=& \text{DReLU}(x-y)\cdot x + (1-\text{DReLU}(x-y))\cdot y\\
=& \text{DReLU}(x-y)\cdot (x-y) +  y = \text{ReLU}(x-y) +  y
\end{aligned}
\end{equation}

\begin{equation} \label{eq:min}
\begin{aligned}
& \text{MIN2}(x,y) = \left\{ 
\begin{aligned}
	& y  \quad x\geq y \quad \\
	& x  \quad x<y \quad \\
\end{aligned}
\right . \\
=& \text{CMP}(x,y)\cdot y + (1-\text{CMP}(x,y))\cdot x\\
=& x - \text{ReLU}(x-y) = \text{ReLU}(y-x) + x
\end{aligned}
\end{equation}
Funnel ReLU (Eq.~\ref{eq:funnelrelu}) is extended from MAX2. 
\begin{equation} \label{eq:funnelrelu}
\begin{aligned}
& \text{Funnel ReLU}(x) = \text{MAX2}(x, \text{T}(x))\\
&(\text{T}(x) \text{is a linear function.}) \\
=& \text{ReLU}(x-\text{T}(x)) +  \text{T}(x)\\
\end{aligned}
\end{equation}

When the number of inputs is more than two, we could compute the MAX, MIN, sorting (SORT), and median (MED) functions. The typical Maxpool function in machine learning is to determine the maximum element from several inputs, e.g., four or nine.

\section{Proofs} \label{app:proofs}
We prove Theorem~\ref{thm:pattern} by the following lemmas. We first prove Lemma~\ref{lmm:truncation} which is stated in~\cite[Sect. 4.1]{MZ17} but not proved. Lemma~\ref{lmm:truncation} explains the one-bit error phenomenon, where Lemma~\ref{lmm:truncation2} presents a special case of Lemma~\ref{lmm:truncation}, i.e., when errors do occur. Lemma~\ref{lmm:oneexist1} further proves the inevitable existence $1$ or $q-1$, which implies Lemma~\ref{lmm:oneexist2}. Lemma~\ref{lmm:oneexist2} exhibits the maximum truncation length required to output $1$ or $q-1$. Lemma~\ref{lmm:tail1} and Lemma~\ref{lmm:tail2} prove the pattern of the tail elements in the truncation result array. Finally, the proofs of these lemmas imply Theorem~\ref{thm:pattern}, i.e., the pattern of the truncation result array. 

In the following, we define $\xi=\xi(x):=x$ if the input $x\in[0,2^{\ell_x})$ and $\xi=\xi(x):=q - x$ if $x\in (q - 2^{\ell_x},q)$, i.e., $\xi$ stays in $[0,2^{\ell_x})$. For an $x$ in $\mathbb{Z}_q=[0,q-1]$, the bit-length of $q$ and $x$ is $\ell$ and $\ell_x$, respectively ($\ell>\ell_x$). The binary form of $\xi$ is defined as $\{\xi_{\ell_x-1},\xi_{\ell_x-2},\cdots,\xi_{1},\xi_{0}\}$, in which $\xi_i$ denotes the $i$-th bit and $\xi:=\sum_{i=0} ^{\ell_x-1} \xi_i\cdot 2^i$. $\lambda$ is the effective bit length of $\xi$, i.e., $\xi_{\lambda-1} = 1$ and $\lambda + 1 < \ell$. The shares of $x$ are $[x]_0$ and $[x]_1$ for $P_0$ and $P_1$, respectively, in which $[x]_0 = r$, $[x]_1 = x - [x]_0$ and $r$ is a random number from $\mathbb{Z}_q$. Moreover, we define 
\begin{align*}
    & \frac{[x]_0}{2^k}:= \text{rShift}([x]_0,k) = [\text{TRC}(x,k)]_0,\\
    & \frac{q - [x]_0}{2^k}:=\text{rShift}(q - [x]_1,k),\\
    & [\text{TRC}(x,k)]_1 = q - \text{rShift}(q - [x]_1, k),
\end{align*}
\begin{align*}
  r:=&r^{\prime\prime}\cdot2^k + r^\prime,\ r^{\prime\prime}\in[0, 2^{\ell-k}),\ r^\prime\in[0, 2^k),\\
  \xi:=&\xi^{\prime\prime}\cdot 2^k + \xi^\prime,\ \xi^{\prime\prime}\in[0, 2^{\ell-k}),\ \xi^\prime\in[0, 2^k),
\end{align*}
then $\frac{\xi}{2^k} = \xi^{\prime\prime}$, for Lemma~\ref{lmm:truncation} and Lemma~\ref{lmm:truncation2}. 
\\
\newline
\noindent \textbf{The proof of Lemma~\ref{lmm:truncation}}:
\begin{IEEEproof}   
  \textbf{Case I}: If $x\in [0,2^{\ell_x})$, then $\xi = x$. Since $[x]_0 = r$, $[x]_1 = x - [x]_0$ and $r\in\mathbb{Z}_q$, the probability of $r \geq \xi$ is $\frac{q - {\xi}}{q}$, which is larger than $\frac{q - 2^{\ell_x}}{q} = 1 - \frac{2^{\ell_x}}{q} > 1 - \frac{2^{\ell_x}}{2^{\ell-1}} = 1 - \frac{1}{2^{\ell - \ell_{\xi} - 1}} = 1 - 2^{\ell_x + 1 - \ell}$. Hence, we will discuss this case under the condition $r\geq\xi$. Let $\textsf{ret} := r - \xi = \textsf{ret}^{\prime\prime}\cdot2^k + \textsf{ret}^\prime$, where $\textsf{ret}^{\prime\prime}\in[0, 2^{\ell-k})$ and $\textsf{ret}^\prime\in[0, 2^k)$. Then, $\frac{\textsf{ret}}{2^k} = \textsf{ret}^{\prime\prime}$. Moreover, $r - \xi = r^{\prime\prime}\cdot2^k + r^\prime - \xi^{\prime\prime}\cdot2^k - \xi^\prime = r^{\prime\prime}\cdot2^k - \xi^{\prime\prime}\cdot2^k + r^\prime - \xi^\prime = (r^{\prime\prime} - \xi^{\prime\prime})\cdot2^k + (r^\prime - \xi^\prime)$. Next, we will prove that $\textsf{ret}^{\prime\prime} = r^{\prime\prime} - \xi^{\prime\prime} - \textsf{bit}$, where $\textsf{bit} = 0$ or $1$. 
  \begin{itemize}
    \item For $r^\prime\geq\xi^\prime$: Due to $r^\prime\in[0, 2^k)$ and $\xi^\prime\in[0, 2^k)$, $r^\prime - \xi^\prime\in[0, 2^k)$. Since $r\geq \xi$, $r^{\prime\prime}\geq \xi^{\prime\prime}$. Then, $r - \xi = r^{\prime\prime}\cdot2^k - \xi^{\prime\prime}\cdot2^k + r^\prime - \xi^\prime = (r^{\prime\prime} - \xi^{\prime\prime})\cdot 2^k + (r^\prime - \xi^\prime)$ satisfying $r^{\prime\prime} - \xi^{\prime\prime}\in[0, 2^{\ell-k})$ and $r^\prime - \xi^\prime\in[0, 2^k)$. Therefore, $\textsf{ret}^{\prime\prime} = (r - \xi)/2^k = r^{\prime\prime} - \xi^{\prime\prime}$. 
    \item For $r^\prime<\xi^\prime$: Due to $r^\prime\in[0, 2^k)$ and $\xi^\prime\in[0, 2^k)$, $2^k + r^\prime - \xi^\prime\in[0, 2^k)$. Since $r\geq \xi$, $r^{\prime\prime}\geq \xi^{\prime\prime} + 1$. Then, $r - \xi = r^{\prime\prime}\cdot2^k - \xi^{\prime\prime}\cdot2^k + r^\prime - \xi^\prime = (r^{\prime\prime} - \xi^{\prime\prime} - 1)\cdot 2^k + (2^k + r^\prime - \xi^\prime)$ satisfying $r^{\prime\prime} - \xi^{\prime\prime} - 1\in[0, 2^{\ell-k})$ and $2^k + r^\prime - \xi^\prime\in[0, 2^k)$. Therefore, $\textsf{ret}^{\prime\prime} = (r - \xi)/2^k = r^{\prime\prime} - \xi^{\prime\prime} - 1$. 
  \end{itemize}
  Hence, $\frac{r - \xi}{2^k} = \textsf{ret}^{\prime\prime} = r^{\prime\prime} - \xi^{\prime\prime} - \textsf{bit}$, where $\textsf{bit} = 0$ or $1$, with a probability larger than $1 - 2^{\ell_x +1-\ell}$. Since $[\text{TRC}(x,k)]_0 = \frac{[x]_0}{2^k} = \frac{r}{2^k} = r^{\prime\prime}$ and $[\text{TRC}(x,k)]_1 = q - \frac{q - [x]_1}{2^k} = q - \frac{r- \xi}{2^k}= q - (r^{\prime\prime} - \xi^{\prime\prime} - \textsf{bit})$, $\text{TRC}(x, k)= [\text{TRC}(x,k)]_0 + [\text{TRC}(x,k)]_1\ \text{mod}\ q= r^{\prime\prime} + q - (r^{\prime\prime} - \xi^{\prime\prime} - \textsf{bit})\ \text{mod}\ q= q + \xi^{\prime\prime} + \textsf{bit}\ \text{mod}\ q = \xi^{\prime\prime} + \textsf{bit}\ \text{mod}\ q = \frac{\xi}{2^k} + \textsf{bit}, \textsf{bit} = 0$ or $1$ with a probability larger than $1 - 2^{\ell_x + 1 - \ell}$. 
  
  \textbf{Case II}: If $x\in (q - 2^{\ell_x},q)$, then $\xi = q - x$. Since $[x]_0 = r$, $[x]_1 = x - [x]_0$ and $r\in\mathbb{Z}_q$, the probability of $\xi + r < q$ is $\frac{q - {\xi}}{q}$, which is larger than $\frac{q - 2^{\ell_x}}{q} = 1 - \frac{2^{\ell_x}}{q} > 1 - \frac{2^{\ell_x}}{2^{\ell-1}} = 1 - \frac{1}{2^{\ell - \ell_{\xi} - 1}} = 1 - 2^{\ell_x + 1 - \ell}$. Hence, we will discuss this case under $\xi + r < q$. Let $\textsf{ret} := r + \xi = \textsf{ret}^{\prime\prime}\cdot2^k + \textsf{ret}^\prime$, where $\textsf{ret}^{\prime\prime}\in[0, 2^{\ell-k})$ and $\textsf{ret}^\prime\in[0, 2^k)$. Then, $\frac{\textsf{ret}}{2^k} = \textsf{ret}^{\prime\prime}$. Moreover, $r + \xi = r^{\prime\prime}\cdot2^k + r^\prime + \xi^{\prime\prime}\cdot2^k + \xi^\prime = r^{\prime\prime}\cdot2^k + \xi^{\prime\prime}\cdot2^k + r^\prime + \xi^\prime = (r^{\prime\prime} + \xi^{\prime\prime})\cdot2^k + (r^\prime + \xi^\prime)$. Next, we will prove that $\textsf{ret}^{\prime\prime} = r^{\prime\prime} + \xi^{\prime\prime} + \textsf{bit}$, where $\textsf{bit} = 0$ or $1$. 
  \begin{itemize}
    \item For $r^\prime + \xi^{\prime}<2^k$: Since $\xi + r< q$, $r + \xi = (r^{\prime\prime} + \xi^{\prime\prime})\cdot2^k + (r^\prime + \xi^\prime)$, and $r^{\prime} + \xi^{\prime}\in[0, 2^k)$, we have $r^{\prime\prime} + \xi^{\prime\prime}\in [0, 2^{\ell-k})$. Then, $r + \xi = r^{\prime\prime}\cdot2^k + \xi^{\prime\prime}\cdot2^k + r^\prime + \xi^\prime = (r^{\prime\prime} + \xi^{\prime\prime})\cdot 2^k + (r^\prime + \xi^\prime)$ satisfying $r^{\prime\prime} + \xi^{\prime\prime}\in[0, 2^{\ell-k})$ and $r^\prime + \xi^\prime\in[0, 2^k)$. Therefore, $\textsf{ret}^{\prime\prime} = (r + \xi)/2^k = r^{\prime\prime} + \xi^{\prime\prime}$. 
    \item For $r^\prime + \xi^{\prime}\geq2^k$: Due to $r^\prime\in[0, 2^k)$ and $\xi^\prime\in[0, 2^k)$, $r^{\prime} + \xi^{\prime} - 2^k \in[0, 2^k)$. Since $\xi + r< q$ and $r + \xi = (r^{\prime\prime} + \xi^{\prime\prime})\cdot2^k + (r^\prime + \xi^\prime) = (r^{\prime\prime} + \xi^{\prime\prime}+1)\cdot2^k + (r^\prime + \xi^\prime - 2^k)$, we have $r^{\prime\prime} + \xi^{\prime\prime} + 1\in [0, 2^{\ell-k})$. Then, $r + \xi = (r^{\prime\prime} + \xi^{\prime\prime} + 1)\cdot 2^k + (r^\prime + \xi^\prime - 2^k)$ satisfying $r^{\prime\prime} + \xi^{\prime\prime} + 1\in[0, 2^{\ell-k})$ and $r^\prime + \xi^\prime - 2^k\in[0, 2^k)$. Therefore, $\textsf{ret}^{\prime\prime} = (r + \xi)/2^k = r^{\prime\prime} + \xi^{\prime\prime} + 1$. 
  \end{itemize}
  Hence, $\frac{r + \xi}{2^k} = \textsf{ret}^{\prime\prime} = r^{\prime\prime} + \xi^{\prime\prime} + \textsf{bit}$, where $\textsf{bit} = 0$ or $1$, with a probability larger than $1 - 2^{\ell_x +1-\ell}$. Since $[\text{TRC}(x,k)]_0 = \frac{[x]_0}{2^k} = \frac{r}{2^k} = r^{\prime\prime}$ and $[\text{TRC}(x,k)]_1 = q - \frac{q - [x]_1}{2^k} = q - \frac{q - (x - r)}{2^k}= q - \frac{q - (q - \xi - r)}{2^k}= q - \frac{\xi + r}{2^k} = q - (r^{\prime\prime} - \xi^{\prime\prime} - \textsf{bit})$, then $\text{TRC}(x, k)= [\text{TRC}(x,k)]_0 + [\text{TRC}(x,k)]_1\ \text{mod}\ q= r^{\prime\prime} + q - (r^{\prime\prime} + \xi^{\prime\prime} + \textsf{bit})\ \text{mod}\ q= q - \xi^{\prime\prime} - \textsf{bit} = q - \frac{\xi}{2^k} - \textsf{bit}, \textsf{bit} = 0$ or $1$ with a probability larger than $1 - 2^{\ell_x + 1 - \ell}$. 
\end{IEEEproof}

Based on Lemma~\ref{lmm:truncation}, Lemma~\ref{lmm:truncation2} describes when the one-bit error would exist.
\begin{lemma} \label{lmm:truncation2}
If $\ell: =\log_2q > \ell_x + 1$, $[x]_0=r$, and $[x]_1=x-[x]_0\ \text{mod}\ q$, the following results hold with a probability of $1 - 2^{\ell_x + 1 - \ell}$: 
\begin{itemize}
  \item If $x = \xi$, then $\text{TRC}(x, k) = \frac{\xi}{2^k}$ for $r^\prime \geq \xi^\prime$, and $\text{TRC}(x, k) = \frac{\xi}{2^k} + 1$ for $r^\prime < \xi^\prime$. 
  \item If $x = q - \xi$, then $\text{TRC}(x, k) = q - \frac{\xi}{2^k}$ for $r^\prime + \xi^\prime < 2^k$, and $\text{TRC}(x,k) = q - \frac{\xi}{2^k} - 1$ for $r^\prime+\xi^\prime \geq 2^k$. 
\end{itemize}
\end{lemma}
Lemma~\ref{lmm:truncation2} can be proved in the similar way as Lemma~\ref{lmm:truncation}. 

We define the following notations $r_{\lambda}^{\prime\prime},r_{\lambda}^{\prime},r_{\lambda-1}^{\prime\prime},r_{\lambda-1}^{\prime},$\\$\xi_{\lambda}^{\prime\prime},\xi_{\lambda}^{\prime},\xi_{\lambda-1}^{\prime\prime},\xi_{\lambda-1}^{\prime}$ for Lemma~\ref{lmm:oneexist1} and Lemma~\ref{lmm:oneexist2}.
\begin{align*}
  r:=&r_{\lambda}^{\prime\prime} \cdot 2^{\lambda} + r_{\lambda} ^\prime,\ r_{\lambda}^{\prime\prime} \in [0, 2^{\ell-\lambda}),\ r^\prime_{\lambda} \in [0, 2^{\lambda}),\\
  r:=&r_{\lambda-1}^{\prime\prime} \cdot 2^{\lambda-1} + r_{\lambda-1}^\prime,\ r_{\lambda-1}^{\prime\prime}\in[0, 2^{l-\lambda+1}),\\
  &r^\prime \in[0, 2^{\lambda-1}),\\
  \xi:=&\xi_{\lambda}^{\prime\prime} \cdot 2^{\lambda} + \xi_{\lambda}^\prime,\ \xi_{\lambda}^{\prime\prime} \in[0, 2^{\ell-\lambda}),\ \xi_{\lambda}^\prime \in[0, 2^{\lambda}),\\
  \xi:=&\xi_{\lambda-1}^{\prime\prime} \cdot 2^{\lambda-1} + \xi_{\lambda-1}^\prime,\ \xi_{\lambda-1}^{\prime\prime} \in [0, 2^{\ell-\lambda+1}),\\
  &\xi_{\lambda-1}^\prime \in[0, 2^{\lambda-1}).
\end{align*}

\begin{figure*}[t!]
  \centering
  \small
  \begin{tikzpicture}[x=1cm,y=1cm,cap=round,align=center,
      fact/.style={rectangle, draw, rounded corners=1mm, fill=white, drop shadow,
            text centered, anchor=center, text=black},growth parent anchor=center,
      fact2/.style={rectangle, draw, rounded corners=1mm, fill=white,
            text centered, anchor=center, text=black},growth parent anchor=center]
  
  
  
    \node[fact, rectangle, draw = black, minimum height = 5.5cm, minimum width = 13.5cm, anchor = west] at (-9, -3.25) {}; 
      
    \node[fact, rectangle, draw = black, minimum height = 5.5cm, minimum width = 2cm,    anchor = west] at (7, -3.25) {}; 
  
      \node at (-9,-0.25) [draw=none, anchor=west] {\small Input: $x = 0b0010110 = 22$; the ring modulus $q=2^{16}=65,536$; random bit $t$ = 1 };

      \node at (7.45,-0.75) [draw=none, anchor=center] {\small $P_2$};
      \node at (-8.75,-0.75) [draw=none, anchor=west] {\small $P_0$, $P_1$ (global view); set $[x]:={(-1)}^t\cdot[x]$, i.e., $x=q-22=65,514$};
      \node at (-8.75,-5.75) [draw=none, anchor=west] {\small $P_0$ and $P_1$ set the shares $[t\oplus \text{DReLu}(x)^\prime] = [1]$};

      \node (t1) at (-9, -3.25) [anchor=west] {
        \footnotesize
        \renewcommand\arraystretch{1.2}
        \begin{tabular}{|l|l|}
          \hline 
          Opt.     & Value \\
          \hline 
          $u_{*}:= {(-1)}^t$     & 65,535 \\
          \hline 
          $u_{0}:= x$            & 65,514\\ 
          \hline 
          $u_1:=\text{TRC}(x,1)$ & 65,525\\ 
          \hline 
          $u_2:=\text{TRC}(x,2)$ & 65,530\\
          \hline 
          $u_3:=\text{TRC}(x,3)$ & 65,534\\
          \hline 
          $u_4:=\text{TRC}(x,4)$ & 65,535\\
          \hline 
          $u_5:=\text{TRC}(x,5)$ & 65,535\\
          \hline 
          $u_6:=\text{TRC}(x,6)$ & 65,535\\
          \hline 
          $u_7:=\text{TRC}(x,7)$ & 0\\
          \hline 
          $u_8:=\text{TRC}(x,8)$ & 0\\
          \hline 
        \end{tabular}
      };

      \node (t2) at (-5, -3.25) [anchor=west] {
        \footnotesize
        \renewcommand\arraystretch{1.2}
        \begin{tabular}{|l|l|}
          \hline 
          Opt.     & Value \\
          \hline 
          $v_{*}:= u_{*}+3\cdot u_0 - 1$            & 65,468 \\
          \hline 
          $v_0:= (\sum_{k=0}^8 u_k) - 1$            & 65,491\\
          \hline 
          $v_1:= (\sum_{k=1}^8 u_k) - 1$            & 65,513\\
          \hline 
          $v_2:= (\sum_{k=2}^8 u_k) - 1$            & 65,524\\
          \hline 
          $v_3:= (\sum_{k=3}^8 u_k) - 1$            & 65,530\\
          \hline 
          $v_4:= (\sum_{k=4}^8 u_k) - 1$            & 65,532\\
          \hline 
          $v_5:= (\sum_{k=5}^8 u_k) - 1$            & 65,533\\
          \hline 
          $v_6:= (\sum_{k=6}^8 u_k) - 1$            & 65,534\\
          \hline 
          $v_7:= (\sum_{k=7}^8 u_k) - 1$            & 65,535\\
          \hline 
          $v_8:= (\sum_{k=8}^8 u_k) - 1$            & 65,535\\
          \hline 
        \end{tabular}
      };

      \node (t3) at (-0.3, -3.25) [anchor=west] {
        \footnotesize
        \renewcommand\arraystretch{1.2}
        \begin{tabular}{|l|l|}
          \hline 
          Opt.     & Value \\
          \hline 
          $w_*:= \Pi\{r_i \cdot v_i\}_*$     & $65,535 r_7$ \\
          \hline 
          $w_0:= \Pi\{r_i \cdot v_i\}_0$     & $65,534 r_6$ \\
          \hline 
          $w_1:= \Pi\{r_i \cdot v_i\}_1$     & $65,491 r_0$ \\
          \hline 
          $w_2:= \Pi\{r_i \cdot v_i\}_2$     & $65,513 r_1$ \\
          \hline 
          $w_3:= \Pi\{r_i \cdot v_i\}_3$     & $65,535 r_8$ \\
          \hline 
          $w_4:= \Pi\{r_i \cdot v_i\}_4$     & $65,533 r_5$ \\
          \hline 
          $w_5:= \Pi\{r_i \cdot v_i\}_5$     & $65,468 r_*$ \\
          \hline 
          $w_6:= \Pi\{r_i \cdot v_i\}_6$     & $65,532 r_4$ \\
          \hline 
          $w_7:= \Pi\{r_i \cdot v_i\}_7$     & $65,530 r_3$ \\
          \hline 
          $w_8:= \Pi\{r_i \cdot v_i\}_8$     & $65,524 r_2$ \\
          \hline 
        \end{tabular}
      };

      \draw [->] (-5.2,-3.25) -- (-4.8,-3.25);
      \draw [->] (-0.5,-3.25) -- (-0.1,-3.25);

      \node at (5.75,-0.75) [draw=none, anchor=center] {\small $[\Pi\{w_k\}]$};
      \draw [->] (4.5,-1) -- (7,-1);
  
      \draw [->] (7,-5) -- (4.5,-5);
      \node at (5.75,-5.25) [draw=none, anchor=center] {\small $[0]$};
  
      \node at (7.2,-2) [draw=none, anchor=west] {\small There is };
      \node at (7.2,-2.5) [draw=none, anchor=west] {\small not a };
      \node at (7.2,-3) [draw=none, anchor=west] {\small 0 value };
      \node at (7.2,-3.5) [draw=none, anchor=west] {\small in $\{w_i\}$.};
  \end{tikzpicture}

  \caption{DReLU example for a positive input. (Opt.: operation, omitting the module operations. Val.: value. )}
  \label{fig:dreluexp}
  \end{figure*}
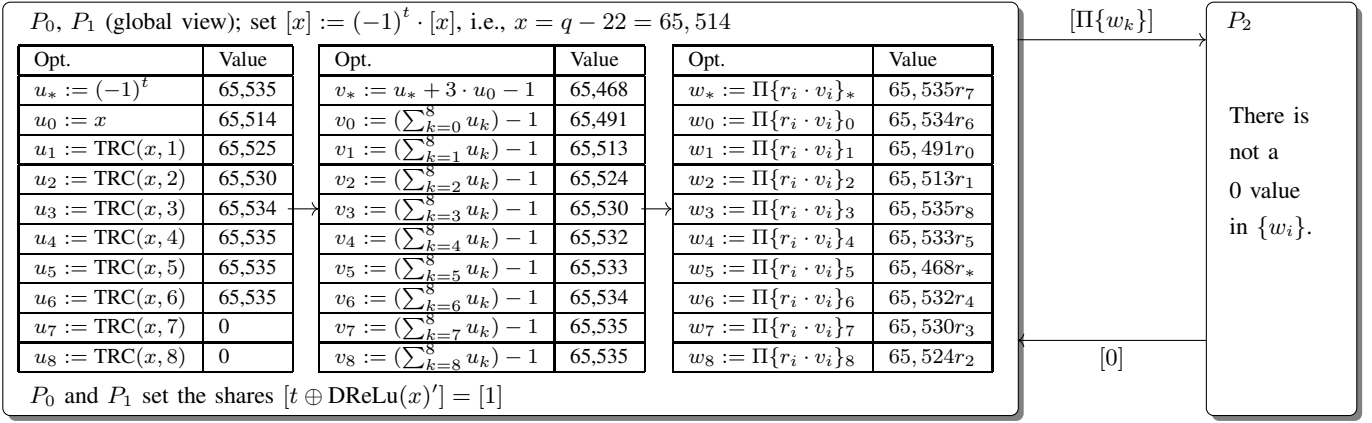
  
\begin{lemma} \label{lmm:oneexist1}
Let $\lambda$ be the effective bit length of $\xi$, i.e., $\xi_{\lambda-1}=1$. Let $\eta$ be the $\lambda$-th significant bit of $r$, i.e., $\eta:=r_{\lambda-1}$. If $\lambda + 1 < \ell$, the following results hold with a probability of $1 - 2^{\lambda + 1 - \ell}$: 
  \begin{itemize}
      \item If $x = \xi$, then $\text{TRC}(x, \lambda - 1) = 1$ or $\text{TRC}(x,\lambda) = 1$.
      \item If $x = q-\xi$, then $\text{TRC}(x,\lambda - 1) = q - 1$ or $\text{TRC}(x,\lambda) = q - 1$.
  \end{itemize}
\end{lemma}
\begin{IEEEproof}
  \textbf{Case I}: For $x = \xi$, Lemma~\ref{lmm:truncation} implies that $\text{TRC}(\xi, \lambda-1) = \frac{\xi}{2^{\lambda-1}} + \textsf{bit} = \xi_{\lambda-1} + \textsf{bit} = 1 + \textsf{bit}$, $\textsf{bit} = 0$ or $1$ with a probability of $1 - 2^ {\lambda + 1 - \ell}$. If $\textsf{bit} = 0$, then $\text{TRC}(\xi, \lambda-1) = 1$. If $\textsf{bit} = 1$, Lemma~\ref{lmm:truncation2} implies $r_{\lambda - 1}^\prime < \xi_{\lambda - 1} ^\prime$. Since $\xi_{\lambda - 1} = 1$, $r_{\lambda}^\prime = \eta\cdot 2^{\lambda-1} + r_{\lambda - 1}^\prime$ and $\xi_{\lambda}^\prime = \xi_{\lambda-1}\cdot 2^{\lambda-1} + \xi_{\lambda - 1}^\prime=2^{\lambda-1} + \xi_{\lambda - 1}^\prime$. Then, $r_{\lambda}^\prime < x_{\lambda}^\prime$. Lemma~\ref{lmm:truncation2} implies $\text{TRC}(\xi,\lambda) = \frac{\xi}{2^\lambda} + 1 = 0 + 1 = 1$. 
  
  \textbf{Case II}: For $x = q - \xi$, Lemma~\ref{lmm:truncation} implies that $\text{TRC}(q - \xi, \lambda - 1) = q - \frac{\xi}{2^{\lambda-1}} - \textsf{bit} = q - \xi_{\lambda-1} - \textsf{bit} = q - 1 - \textsf{bit}$, $\textsf{bit} = 0$ or $1$ with a probability of $1 - 2^{\lambda +1-\ell}$. If $\textsf{bit} = 0$, $\text{TRC}(q - \xi,\lambda-1) = q - 1$. If $\textsf{bit} = 1$, Lemma~\ref{lmm:truncation2} implies $r_{\lambda-1}^\prime + \xi_{\lambda-1}^\prime \geq 2^{\lambda-1}$. Therefore, $r_{\lambda}^\prime = \eta\cdot 2^{\lambda-1} + r_{\lambda-1}^\prime$ and $\xi_{\lambda}^\prime = \xi_{\lambda-1}\cdot 2^{\lambda-1} + \xi_{\lambda-1}^\prime = 2^{\lambda-1} + \xi_{\lambda-1}^\prime$. Then, $r_{\lambda}^\prime + \xi_{\lambda}^\prime = r_{\lambda-1}^\prime + \xi_{\lambda-1}^\prime + 2^{\lambda-1} \geq 2^\lambda$ due to $r_{\lambda-1}^\prime + \xi_{\lambda-1}^\prime \geq 2^{\lambda-1}$. Lemma~\ref{lmm:truncation2} implies $\text{TRC}(q - \xi,\lambda) = q - \frac{\xi}{2^\lambda} - 1 = q - 0 - 1 = q - 1$. 
\end{IEEEproof}

\begin{lemma} \label{lmm:oneexist2}
If $\lambda + 1 < \ell$, for any number $\hat{\ell} \geq \lambda$, the following results hold with a probability of $1 - 2^{\lambda + 1 - \ell}$:
\begin{itemize}
    \item There exists a number $ \hat{\lambda} \leq \hat{\ell}$, satisfying $\text{TRC}(\xi, \hat{\lambda}) = 1$. 
    \item There exists a number $ \hat{\lambda} \leq \hat{\ell}$, satisfying $\text{TRC}(\xi, \hat{\lambda}) = q - 1$. 
\end{itemize}
\end{lemma}
Lemma~\ref{lmm:oneexist2} could be proved via the proof of Lemma~\ref{lmm:oneexist1}. 

Considering any number $\lambda^\prime > \ell_x$, we extra define the following notations $r_{\lambda^\prime}^{\prime\prime},r_{\lambda^\prime}^{\prime},r_{\lambda^\prime-1}^{\prime\prime},r_{\lambda^\prime-1}^{\prime},\xi_{\lambda^\prime}^{\prime\prime},\xi_{\lambda^\prime}^{\prime},\xi_{\lambda^\prime-1}^{\prime\prime},$\\$\xi_{\lambda^\prime-1}^{\prime}$ for Lemma~\ref{lmm:tail1} and Lemma~\ref{lmm:tail2}.
\begin{align*}
  r:=&r_{\lambda^\prime}^{\prime\prime} \cdot 2^{\lambda^\prime} + r_{\lambda^\prime} ^\prime,\ r_{\lambda^\prime}^{\prime\prime} \in [0, 2^{\ell-\lambda^\prime}),\ r^\prime_{\lambda^\prime} \in [0, 2^{\lambda^\prime}),\\
  r:=&r_{\lambda^\prime-1}^{\prime\prime} \cdot 2^{\lambda^\prime-1} + r_{\lambda^\prime-1}^\prime,\ r_{\lambda^\prime-1}^{\prime\prime}\in[0, 2^{l-\lambda^\prime+1}),\\
  &r^\prime \in[0, 2^{\lambda^\prime-1}),\\
  \xi:=&\xi_{\lambda^\prime}^{\prime\prime} \cdot 2^{\lambda^\prime} + \xi_{\lambda^\prime}^\prime,\ \xi_{\lambda^\prime}^{\prime\prime} \in[0, 2^{\ell-\lambda^\prime}),\ \xi_{\lambda^\prime}^\prime \in[0, 2^{\lambda^\prime}),\\
  \xi:=&\xi_{\lambda^\prime-1}^{\prime\prime} \cdot 2^{\lambda^\prime-1} + \xi_{\lambda^\prime-1}^\prime,\ \xi_{\lambda^\prime-1}^{\prime\prime} \in [0, 2^{\ell-\lambda^\prime+1}),\\
  &\xi_{\lambda^\prime-1}^\prime \in[0, 2^{\lambda^\prime-1}).
\end{align*}
\begin{lemma} \label{lmm:tail1}
If $\ell > \lambda + 1$, for any number $\lambda^\prime > \ell_x$, the following results hold with a probability of $1 - 2^{\lambda^\prime - \ell}$:
  \begin{itemize}
      \item If $\text{TRC}(\xi, \lambda^\prime - 1) = 1$, then $\text{TRC}(\xi, \lambda^\prime) = 1$ or $0$.
      \item If $\text{TRC}(q - \xi, \lambda^\prime - 1) = q - 1$, then $\text{TRC}(q - \xi, \lambda^\prime) = q - 1$ or $0$.
  \end{itemize}
\end{lemma}
\begin{IEEEproof}
  \textbf{Case I}: For $\text{TRC}(\xi, \lambda^\prime-1) = 1$, we have $\text{TRC}(\xi, \lambda^\prime-1) = \frac{\xi}{2^{\lambda^\prime-1}} + \textsf{bit}_{\lambda^\prime-1}$ according to Lemma~\ref{lmm:truncation2}, in which $\textsf{bit}_{\lambda^\prime-1}= 1$ or 0, with a probability of $1 - 2^{\lambda^\prime - \ell}$. $\text{TRC}(\xi, \lambda^\prime-1) =1$ implies two sub-cases.
  \begin{itemize}
    \item For $\frac{\xi}{2^{\lambda^\prime-1}} = 1$ and $\textsf{bit}_{\lambda^\prime-1} = 0$, $\frac{\xi}{2^{\lambda^\prime}} = 0$. Since $\frac{\xi}{2^{\lambda^\prime-1}} = 1$, we have $\text{TRC}(\xi, \lambda^\prime) = \frac{\xi}{2^{\lambda^\prime}} + \textsf{bit}_{\lambda^\prime} = \textsf{bit}_{\lambda^\prime}$, where $\textsf{bit}_{\lambda^\prime} = 0$ or $1$. Hence, $\text{TRC}(\xi, \lambda^\prime) = 0$ or $1$.
    \item For $\frac{\xi}{2^{\lambda^\prime-1}} = 0$ and $\textsf{bit}_{\lambda^\prime-1} = 1$, $\frac{\xi}{2^{\lambda^\prime}} = 0$. Since $\frac{\xi}{2^{\lambda^\prime-1}} = 0$, we have $\text{TRC}(\xi, \lambda^\prime) = \frac{\xi}{2^{\lambda^\prime}} + \textsf{bit}_{\lambda^\prime} = \textsf{bit}_{\lambda^\prime}$, where $\textsf{bit}_{\lambda^\prime} = 0$ or $1$. Hence, $\text{TRC}(\xi, \lambda^\prime) = 0$ or $1$.
  \end{itemize}
  
  \textbf{Case II}:	For $\text{TRC}(q - \xi, \lambda^\prime-1) = q - 1$, we have $\text{TRC}(q - \xi, \lambda^\prime-1) = q - \frac{\xi}{2^{\lambda^\prime-1}} - \textsf{bit}_{\lambda^\prime-1}$, with a probability of $1 - 2^{\lambda^\prime - \ell}$, according to Lemma~\ref{lmm:truncation2}.  $\text{TRC}(q - \xi, \lambda^\prime-1) =q - 1$ implies two sub-cases, i.e., $\frac{\xi}{2^{\lambda^\prime-1}} = 1$ and $\textsf{bit}_{\lambda^\prime-1} = 0$, and $\frac{\xi}{2^{\lambda^\prime-1}} = 0$ and $\textsf{bit}_{\lambda^\prime-1} = 1$. Similar to the proof of the two sub-cases of Case I, these two sub-cases both satisfy that $\text{TRC}(q - \xi, \lambda^\prime) = q - 1$ or $0$.
\end{IEEEproof}

\begin{lemma} \label{lmm:tail2}
  If $\ell > \lambda + 1$, for any number $\lambda^\prime > \ell_x$, $\text{TRC}(x, \lambda^\prime) = 0$ if $\text{TRC}(x, \lambda^\prime-1) = 0$, with a probability of $1 - 2^{\lambda^\prime - \ell}$.
\end{lemma}
\begin{IEEEproof}
  \textbf{Case I}: For $x = \xi$, $\text{TRC}(x, \lambda^\prime-1) = \frac{\xi}{2^{\lambda^\prime-1}}+ \textsf{bit}_{\lambda^\prime-1}$ according to Lemma~\ref{lmm:truncation2}, in which $\textsf{bit}_{\lambda^\prime-1}$ = $0$ or $1$, with a probability of $1 - 2^{\lambda^\prime - \ell}$. Under this condition, since $\text{TRC}(x, \lambda^\prime-1) = 0$, we have $\frac{\xi}{2^{\lambda^\prime-1}}=0$ and $\textsf{bit}_{\lambda^\prime-1}=0$. Also, Lemma~\ref{lmm:truncation2} implies $\lambda\leq \lambda^\prime-1$ and $r'_{\lambda^\prime-1}\geq \xi'_{\lambda^\prime-1}$. Hence, $r'_{\lambda^\prime} = r_{\lambda^\prime-1}\cdot 2^{\lambda^\prime-1} + r'_{\lambda^\prime-1}$ and $\xi'_{\lambda^\prime} = \xi_{\lambda^\prime-1}\cdot 2^{\lambda^\prime-1} + \xi'_{\lambda^\prime-1} = \xi_{\lambda^\prime-1}$, where $r_{\lambda^\prime-1}$ and $\xi_{\lambda^\prime-1}$ are $(\lambda^\prime-1)$-th bits of $r$ and $\xi$ respectively. Therefore, $r'_{\lambda^\prime}\geq \xi'_{\lambda^\prime}$. Then, $\text{TRC}(\xi, \lambda^\prime) = \frac{\xi}{2^{\lambda^\prime}} + \textsf{bit}_{\lambda^\prime} = 0$, where $\frac{\xi}{2^{\lambda^\prime}} = 0$ and $\textsf{bit}_{\lambda^\prime} = 0$, due to Lemma~\ref{lmm:truncation2}.

  \textbf{Case II}: For $x = q - \xi$, $\text{TRC}(x, \lambda^\prime-1) = q - \frac{\xi}{2^{\lambda^\prime-1}} - \textsf{bit}_{\lambda^\prime-1}$ according to Lemma~\ref{lmm:truncation2}, in which $\textsf{bit}_{\lambda^\prime-1}$ = $0$ or $1$, with a probability of $1 - 2^{\lambda^\prime - \ell}$. Under this condition, since $\text{TRC}(x, \lambda^\prime-1) = 0$, we have $\frac{\xi}{2^{\lambda^\prime-1}}=0$ and $\textsf{bit}_{\lambda^\prime-1}=0$. Also, Lemma~\ref{lmm:truncation2} implies $\lambda\leq \lambda^\prime-1$ and $r'_{\lambda^\prime-1} + \xi'_{\lambda^\prime-1}< 2^{\lambda^\prime-1}$. Hence, $r'_{\lambda^\prime} = r_{\lambda^\prime-1}\cdot 2^{\lambda^\prime-1} + r'_{\lambda^\prime-1}$ and $\xi'_{\lambda^\prime} = \xi_{\lambda^\prime-1}\cdot 2^{\lambda^\prime-1} + \xi'_{\lambda^\prime-1} = \xi'_{\lambda^\prime-1}$, where $r_{\lambda^\prime-1}$ and $\xi_{\lambda^\prime-1}$ are $(\lambda^\prime-1)$-th bits of $r$ and $\xi$ respectively. Therefore, $r'_{\lambda^\prime} + \xi'_{\lambda^\prime}< 2^{\lambda^\prime}$. Then, $\text{TRC}(x, \lambda^\prime) = \text{TRC}(q - \xi, \lambda^\prime)=q - \frac{\xi}{2^{\lambda^\prime}} - \textsf{bit}_{\lambda^\prime} = 0$, where $\frac{\xi}{2^{\lambda^\prime}} = 0$ and $\textsf{bit}_{\lambda^\prime} = 0$, due to Lemma~\ref{lmm:truncation2}.
\end{IEEEproof}

Finally, all the proofs of Lemma~\ref{lmm:truncation}, Lemma~\ref{lmm:truncation2}, Lemma~\ref{lmm:oneexist1}, Lemma~\ref{lmm:oneexist2}, Lemma~\ref{lmm:tail1}, and Lemma~\ref{lmm:tail2} imply Theorem~\ref{thm:pattern}.

\section{DReLU Example} \label{app:dreluexp}
Fig.~\ref{fig:dreluexp} shows a DReLU example for a positive input $x = 0b00010110 = 22$. In this case, we set the ring $\mathbb{Z}_q=[0,65535],q=65536,\ell=16$. If $P_0$ and $P_1$ generate a random bit $t = 1$, they would reverse the sum of their input shares. Then, $u_*$ would be $q-1=65535,u_0=q-x=65,514$. After the repeated times of probabilistic truncation, the $\{u_i\}$ array is obtained. The errors occur at $\text{TRC}(x,2)$, $\text{TRC}(x,5)$, and $\text{TRC}(x,6)$. In order to prevent the repeating $(q-1)$'s in $\{u_i\}$ leaking information about the input's range, $P_0$ and $P_1$ compute the summation of the subarrays of $\{u_i\}$ (excluding $v_*$) and then subtracting one from each summation. Next, $P_0$ and $P_1$ apply random maskings and a random shuffle using $\textsf{seed}_{01}$, leading to the shares of $\{w_i\}$. After reconstructing $\{w_i\}$, if $P_2$ finds out there is no 0's in the array, then the blinded input is negative, and $P_2$ sends the shares of $\text{DReLu}(x)^\prime:=0$ to $P_0$ and $P_1$. Finally, $P_0$ and $P_1$ execute an XOR operation to obtain the shares of the output $\text{DReLu}(x):=\text{DReLu}(x)^\prime\oplus t = 0 \oplus 1 = 1$.

\section{Response to Security Concerns Raised by Xu et al.~\cite{XLH24}} \label{app:fix}
In response to the security analysis of the DReLU protocol within our paper as proposed by Xu et al.~\cite{XLH24}, we offer the following clarification:

In the field of MPC, it is customary to perform a re-share operation on shares before they are reconstructed. This ensures that the shares being reconstructed are uniformly random. Consequently, the participant obtaining the shares cannot glean any information beyond the results intended to be revealed through the reconstructing process.

In the realm of MPC, re-shareing is a standard operation and, as with many other MPC papers, we did not specifically emphasize the re-share step in our protocol. However, this omission does not imply that re-sharing is not required.

The re-share process typically proceeds as follows: Parties $P_0$ and $P_1$ locally generate a one-time random number $r$ from the ring $\mathbb{Z}_n$. Subsequently, $P_0$ and $P_1$ locally compute $[a^\prime]_0=[a]_0+r\ \text{mod}\ n$, and $[a^\prime]_1=[a]_1-r\ \text{mod}\ n$ respectively. This ensures that when $[a^\prime]_0$ and $[a^\prime]_1$ are uniformly random, meaning that they do not carry any information related to prior computations.

Therefore, the attack scenario posited by Xu et al.~\cite{XLH24}, based on ``$P_0$ and $P_1$ directly sending $[w_i]_0=r\cdot[v_i]_0$, $[w_i]_1=r\cdot[v_i]_1$ to $P_2$,'' does not occur in the actual execution of our protocol.

\end{document}